\documentclass[usenatbib]{mnras}

\usepackage{geometry}

\usepackage[T1]{fontenc}
\usepackage[latin9]{inputenc}
\usepackage{amsmath}
\usepackage{amsfonts}
\usepackage{amssymb}
\usepackage{graphicx}
\usepackage{hyperref}
\usepackage{xcolor}           

\title[The baryon fraction in halos as a cause of stellar mass scatter in EAGLE]{The evolution of the baryon fraction in halos as a cause of scatter in the galaxy stellar mass in the EAGLE simulation}
\author[A. Kulier et al.]{Andrea Kulier$^{1}$\thanks{E-mail: \href{mailto:akulier@astro.puc.cl}{akulier@astro.puc.cl}}, Nelson Padilla$^{1}$, 
Joop Schaye$^{2}$, Robert A. Crain$^{3}$, \newauthor Matthieu Schaller$^{2, 4}$, Richard G. Bower$^{4}$, Tom Theuns$^{4}$, Enrique Paillas$^{1}$\\
$^{1}$Instituto de Astrofisica, Pontificia Universidad Cat\'{o}lica de Chile, Av. Vicu\~{n}a Mackenna 4860, Santiago, Chile\\
$^{2}$Leiden Observatory, Leiden University, P.O. Box 9513, NL-2300 RA Leiden, The Netherlands\\
$^{3}$Astrophysics Research Institute, Liverpool John Moores University, 146 Brownlow Hill, Liverpool L3 5RF, UK\\
$^{4}$Institute for Computational Cosmology, Department of Physics, University of Durham, South Road, Durham, DH1 3LE, UK
}

\begin{document}

\date{\today}

\pagerange{\pageref{firstpage}--\pageref{lastpage}} \pubyear{2018}

\maketitle

\label{firstpage}

\begin{abstract}
The EAGLE simulation suite has previously been used to investigate the relationship between
the stellar mass of galaxies, $M_{*}$, and the properties of dark matter halos, using the 
hydrodynamical reference simulation combined with a dark matter only (DMO) simulation
having identical initial conditions.  The stellar 
masses of central galaxies in halos with $M_{\mathrm{200c}} > 10^{11} \mathrm{M_{\odot}}$ 
were shown to correlate with the DMO halo maximum circular velocity, 
with \mbox{$\approx 0.2$} dex of scatter
 that is uncorrelated with other DMO halo properties.
Here we revisit the origin of the scatter in the $M_{*}-V_{\mathrm{max, DMO}}$ relation in EAGLE at $z = 0.1$.
We find that the scatter in $M_{*}$ correlates with the mean age of the galaxy stellar population
such that more massive galaxies at fixed $V_{\mathrm{max, DMO}}$ are younger. The scatter
in the stellar mass and mean stellar population age results from variation in 
the baryonic mass, $M_{\mathrm{bary}} = M_{\mathrm{gas}} + M_{*}$,
of the galaxies' progenitors at
fixed halo mass and concentration. At the redshift of peak correlation ($z \approx 1$), the progenitor baryonic mass 
accounts for $75\%$ of the variance in the $z=0.1$ $M_{*}-V_{\mathrm{max, DMO}}$ relation. 
The scatter in the baryonic mass, in turn, is primarily set by differences in feedback strength and gas accretion
over the course of the evolution of each halo.

\end{abstract}


\section{Introduction}

Understanding the relationship between galaxies and their host dark matter
halos has been a longstanding problem relevant to both galaxy evolution and cosmology. 
Owing to the difficulty of directly measuring the properties of dark matter halos, it is often
necessary to infer them from the observable properties of the galaxies that they host. 
Therefore, relations between measurable galaxy properties and halo properties have been
much sought after.

Hydrodynamical cosmological simulations offer a way to investigate these relationships. 
However, such simulations were until recently unable to produce large enough samples of galaxies
at sufficient resolution to perform statistical studies of galaxy properties.
Partly as a result, a variety of methods have been created for the purpose of assigning
galaxies to dark matter halos from dark matter-only simulations, which are much less computationally
expensive to perform. These include halo occupation
distributions \citep{seljak2000, peacock2000} and abundance matching 
\citep{valeostriker, vale2004, kravtsov2004}. Such models are generally 
calibrated to reproduce the observed properties of populations of galaxies; e.g., their spatial clustering.

In contrast to their predecessors, recent hydrodynamical cosmological simulations
such as EAGLE \citep{eagleschaye, eaglecrain}, Illustris \citep{illustris0, illustris1, illustris2},
and Horizon-AGN \citep{dubois} allow for measurements of galaxy
and halo properties for sizeable galaxy populations. Such simulations can be used
to study galaxy-halo relations and to inform 
semi-analytic methods such as those previously mentioned.

One topic that has recently been investigated with the latest hydrodynamical simulations 
is the correlation between
galaxy stellar masses and the properties of their host dark matter halos.
 This is particularly relevant to abundance matching models, which assign
observed samples of galaxies to simulated dark matter halos by assuming a monotonic 
relation (with some scatter) between galaxy stellar mass or luminosity and a given dark matter halo parameter.
Simulations can be used to identify the most suitable halo property 
by which to assign galaxy stellar masses to halos. 
The EAGLE simulation suite has been used for this purpose because 
 it reproduces the galaxy stellar mass function \citep{eagleschaye}, 
which is reproduced by construction in halo abundance matching models,
and because it includes a dark matter-only variant of the main hydrodynamical simulation 
with identical initial conditions, allowing the identification of ``corresponding'' host dark matter
halos in the dark matter-only simulation.

In particular, \citet{chaves2016} and \citet{matthee2017} both used the set of EAGLE
simulations to examine the relationship between the
stellar mass of galaxies and the properties of their matched dark matter halos in the dark-matter only simulation.
\citet{chaves2016} found that the stellar mass of central and satellite galaxies is most tightly correlated with 
the parameter $V_{\mathrm{relax}}$, the maximum circular velocity attained by the host halo in its history while
satisfying a relaxation criterion. This parameter had slightly less scatter with the stellar mass than $V_{\mathrm{peak}}$,
the maximum circular velocity achieved by the halo during its entire history, and $V_{\mathrm{infall}}$, the maximum 
circular velocity of the halo before it becomes a subhalo of a larger halo.
Furthermore, the authors found that parameters
based on the maximum circular velocity of the halo are more strongly correlated with the galaxy stellar mass than those
based on the halo mass. This is in agreement with results from abundance matching fits
to observed halo clustering (e.g, \citealt{reddick2013}).

\citet{matthee2017} considered only central galaxies, obtaining results consistent with \citet{chaves2016}.
They found that $V_{\mathrm{max}}$ in the dark matter-only simulation 
correlates better with the stellar mass $M_{*}$ than the
halo mass $M_{\mathrm{200c}}$. However, there was a remaining scatter of $\approx 0.2$ dex 
in the correlation between $V_{\mathrm{max}}$ and $M_{*}$ 
for their halo sample, defined by a mass cut of $M_{\mathrm{200c}} > 10^{11} \mathrm{M_{\odot}}$. 
Interestingly, they found that the residuals of the $V_{\mathrm{max}} - M_{*}$ relation
did not correlate with \textit{any} of several other halo parameters that they considered ---
including concentration, half-mass formation time, sphericity, triaxiality, spin, and two simple measures of 
small- and large-scale environment. 

In this paper, we investigate the source of the scatter in the relation between $V_{\mathrm{max}}$
and $M_{*}$ for central galaxies.
In contrast to \citet{chaves2016} and \citet{matthee2017}, we focus on correlations between the scatter and the baryonic properties of galaxies and halos.
In Section 2 we describe the EAGLE simulation suite used in our analysis and how
we selected our sample of halos. In Section 3 we present our results on the
origin of the $V_{\mathrm{max}} - M_{*}$ scatter at $z = 0.1$. Finally, we summarize our conclusions
in Section 4.

Throughout this paper we assume the Planck cosmology \citep{planck}
adopted in the EAGLE simulation,
such that $h = 0.6777$, $\Omega_{\Lambda} = 0.693$, $\Omega_{m} = 0.307$,
and $\Omega_{b} = 0.048$.

\section{Simulations and halo sample}

\subsection{Simulation overview}
\label{sec:simulation}

EAGLE \citep{eagleschaye, eaglecrain, mcalpine} is a suite of cosmological hydrodynamical simulations, run using
a modified version of the N-body smooth particle hydrodynamics (SPH) code GADGET-3 \citep{gadget}.
The changes to the hydrodynamics solver, referred to as ``Anarchy'' and described in \citet{anarchy},
are based on the formulation of SPH in \citet{hopkins}, and include changes to the 
handling of the viscosity \citep{cullen}, the conduction \citep{price}, the smoothing kernel \citep{denhen}, and 
the time-stepping \citep{durier}.

The reference EAGLE simulation has a box size of 100 comoving Mpc per side, containing
$1504^{3}$ particles each of dark matter and baryons, with a dark matter particle
mass of $9.70\times10^{6} \mathrm{M_{\odot}}$, and an initial gas (baryon) particle mass of $1.81\times10^{6} \mathrm{M_{\odot}}$.
The Plummer-equivalent gravitational softening length is 2.66 comoving kpc until $z = 2.8$ and
0.70 proper kpc afterward. The EAGLE suite also includes a second simulation containing 
only dark matter that has the same total cosmic matter density,
resolution, initial conditions,
and number of dark matter particles (each with mass $1.15\times10^{7} \mathrm{M_{\odot}}$)
as the reference simulation. 

Subgrid physics in EAGLE includes radiative cooling, photoionization heating,
star formation, stellar mass loss, stellar feedback, 
supermassive black hole accretion and mergers, and AGN feedback. 
Here we briefly summarize these subgrid prescriptions, which are described
in more detail in \citet{eagleschaye}.

Radiative cooling and photoionization heating is implemented
using the model of \citet{cooling}. Cooling and heating rates are computed for 11 elements
using CLOUDY \citep{cloudy}, assuming that the gas is optically thin, in ionization equilibrium,
and exposed to the cosmic microwave background and the evolving \citet{hardtmadau} UV and X-ray
background that is imposed instantaneously at $z = 11.5$. Extra energy is also
injected at this redshift and at $z = 3.5$ to model HI and HeII reionization
respectively.

Gas particles undergo stochastic conversion into star particles using the prescription
of \citet{sfr}, which imposes the Kennicutt-Schmidt law \citep{ks} on the gas. 
A metallicity-dependent density threshold for gas to become star-forming
 is used based on \citet{sfr2}. Star particles are assumed to be simple 
stellar populations with a \citet{chabrier} initial mass function. 
The prescriptions for stellar evolution and mass loss from \citet{snwiersma} are used.
The fraction of the initial stellar particle mass that is
leaving the main sequence at each time step is used in combination with the initial
elemental abundances of the star particle to compute the mass that
is ejected from the particle due to stellar winds and supernovae.

To model the effect of stellar feedback on the ISM, the stochastic feedback prescription
of \citet{supernova} is used, in which randomly selected gas particles close to a star particle 
that is losing energy are instantly heated by $10^{7.5}$ K.
Each star particle is assumed to lose the total amount of energy produced by type II supernovae
in a Chabrier IMF when it reaches an age of 30 Myr.
The strength of the feedback in EAGLE is calibrated by adjusting the fraction
of this energy that is assumed to heat the nearby gas.

Halos that reach a mass of $10^{10} \mathrm{M_{\odot}}/h$ are seeded with black holes of subgrid mass
$10^{5} \mathrm{M_{\odot}}/h$ at their centers by converting the most bound gas particle
into a ``black hole'' seed particle \citep{agn1}. These particles accrete mass at a rate specified
by the minimum of the Eddington rate and the modified Bondi-Hoyle accretion rate from \citet{agn2} with $\alpha = 1$.
Black hole particles are also able to merge with one another.

AGN feedback is modeled in a stochastic
manner similar to stellar feedback, with the energy injection rate proportional
to the black hole accretion rate. In contrast to the stellar feedback, 
adjustment of the fraction of lost energy assumed to heat the gas does not significantly 
affect the masses of galaxies due to self-regulation \citep{booth2010}.

The feedback scheme used by EAGLE 
is able to approximately reproduce the local 
galaxy stellar mass function;
some differences near the ``knee'' of the distribution cause the EAGLE stellar mass
density to be $\approx 20\%$ lower than that inferred from observations.
The feedback parameters have been calibrated so as to additionally reproduce the distribution 
of present-day galaxy sizes \citep{eaglecrain}. EAGLE has been found to reproduce, without further parameter
calibration, a number of other observed features of the population of galaxies, such as the $z = 0$ Tully-Fisher
relation, specific star formation rates, rotation curves, colors, and
the evolution of the galaxy stellar mass function and galaxy sizes 
\citep{eagleschaye, furlong2015, anarchy, trayford2016, furlong2017}.

\subsection{Halo/galaxy sample and properties}
\label{sec:sample}

Halos in EAGLE are identified by applying a friends-of-friends (FoF)
algorithm with a linking length of $b = 0.2$ times the mean interparticle separation
to the distribution of dark matter particles \citep{davis}. Other particles types (gas, stars, and black holes) are assigned to 
the FoF halo of the nearest dark matter particle. 
The SUBFIND \citep{subfind1, subfind2} algorithm is then used to identify local overdensities of all particles types 
within FoF halos --- referred to as subhalos. SUBFIND assigns to each subhalo only those particles 
that are gravitationally bound to it, with no overlap in particles between distinct
subhalos. When we refer to ``galaxies'', we are referring to the baryonic particles
associated with each subhalo. The
subhalo in each FoF halo that contains the most bound particle is defined to be the central
subhalo, and all others are defined as satellites. The location of the most bound particle is also used to define
the center of the FoF halo, around which mean spherical overdensities are calculated to obtain halo masses such
as $M_{\mathrm{200c}}$, the mass inside the radius within which the mean overdensity is 200 times
the critical density of the Universe. 

The FoF and SUBFIND algorithms are run at a series of 29 simulation snapshots
from $z = 20$ to $z = 0$, with the time between snapshots increasing from
$\approx 0.1$ Gyr at the beginning of the simulation to $\approx 1$ Gyr
at the end. Galaxy and halo catalogs as well as particle data from EAGLE have been made 
publicly available \citep{mcalpine}.

We use the method described in \citet{schaller2015} to match halos from the reference hydrodynamic
simulation to those from the dark matter-only (DMO) simulation, and the reader is referred to that paper
for details. To summarize, the reference and DMO EAGLE simulations have identical
initial conditions save for the fact that the DMO simulation has slightly more massive
dark matter particles to account for the mass in baryons present in the reference simulation.
Each particle is tagged with a unique identifier
where two particles with the same identifier in the two simulations have the same
initial conditions. We define two subhalos in the reference simulation and the DMO simulation to
correspond to one another if they share at least half of their 50 most bound particles.

We take as our primary sample in the reference simulation one identical to that of \citet{matthee2017}:
central galaxies with redshift $z = 0.1$ and host halo mass $M_{\mathrm{200c}} > 10^{11} \mathrm{M_{\odot}}$,
resulting in a sample of 9929 galaxies and their host halos. 
We successfully match 9774 of these halos ($98.4\%$) in the DMO simulation. However,
we discard the halos whose matches in the DMO simulation are satellite subhalos rather than
centrals, leaving 9543 halos ($96.1\%$ of our original sample).

In our analysis, we consider the properties of the progenitors of our galaxy sample in 
order to determine the origin of the scatter in their stellar masses. Merger
trees have been created from the EAGLE simulation snapshots using a modified version \citep{qu} of
the D-TREES algorithm \citep{dtrees}. D-TREES links subhalos to their descendants
by considering the $N_{\mathrm{link}}$ most bound particles and identifying
the subhalo that contains the majority of these particles in the next time snapshot.
For EAGLE, $N_{\mathrm{link}}$ is set to be min$(100, \mathrm{\mathrm{max}}(0.1N_{\mathrm{subhalo}}, 10))$,
where $N_{\mathrm{subhalo}}$ is the total number of particles in the subhalo.
Each subhalo is assigned only a single descendant, but a subhalo may have multiple progenitors.
Each subhalo with at least one progenitor has a single ``main progenitor'', defined
as the progenitor that has the largest mass summed across all earlier outputs,
as suggested by \citet{delucia} to avoid swapping of the main progenitor during major mergers.
In some cases, galaxies can disappear in a snapshot and reappear at a later time;
because of this, descendants are identified up to 5 snapshots later.

Essentially all ($99.9\%$) of the galaxies in our $z = 0.1$ main sample have at least
one progenitor up to $z = 4$, although in this paper we mainly concern ourselves with
$z \le 2$. We investigate the correlations between the properties of the central galaxies/subhalos and
their FoF host halos at $z = 0.1$ and the properties of their progenitors at each prior timestep. 
We do this using the properties of the main progenitor subhalo and its FoF host halo, as well as the combined
properties of all the progenitor subhalos. In the latter case, we consider
subhalos of any mass that have a non-zero mass
in bound stars or gas to be progenitors.

We use as galaxy stellar masses the total stellar mass assigned to
each galaxy's subhalo by SUBFIND, which includes some diffuse stellar mass that is similar to
``intracluster light''. This differs from the definition in \citet{matthee2017}, who
used only the stellar mass within 30 kpc, although they found that their analysis would be 
nearly identical if they had used the total stellar mass because the two masses are only significantly
different in very massive halos. 

As a measure of the age of each galaxy's stellar population, we use 
the initial-mass-weighted mean stellar age. This is the mean age of the
star particles belonging to a galaxy weighted by their initial mass---the mass of each star
particle at the moment it formed from a gas particle, before it has lost mass
due to stellar winds and supernovae (see \S\hyperref[sec:simulation]{2.1}).

We also examine the baryonic masses (stars and gas) within halos in EAGLE. 
(We do not include black hole particles, as they are a minuscule fraction of the
total baryonic mass in each halo.) For each subhalo, 
we take the sum of the masses of the bound stellar and
gas particles, including both hot and cold phase gas, to be its 
baryonic mass.

When analyzing the main progenitors of our central
galaxy sample, we consider all the baryonic mass within the host
FoF halo to be potentially collapsing onto
the central galaxy. We define the total baryonic mass of an FoF halo as
the sum of the baryonic masses of all its subhalos; this means
that we only include the gas mass that is gravitationally bound
to substructures within the FoF halo. For the highest-mass halos in our sample,
gas that is bound to substructure constitutes nearly all of the gas mass,
but the variance in the ratio of unbound to bound gas 
increases significantly with decreasing halo mass, 
such that low-mass halos in our sample may contain less bound than unbound gas mass.

The above analysis is complicated by the fact that 
the main progenitor of a central galaxy/subhalo is not always a central. 
There are two possible causes for this. 
One is that a central subhalo can interact with
a satellite, and mass exchange between
the two can cause the satellite
to become the new central subhalo of the
FoF halo (defined as containing the most bound particle).
Such interactions can happen during the course of a merger,
and if the two subhalos merge to become a new central,
its main progenitor (defined as the one with the most massive
total mass history) may be a satellite during some snapshots.

The second cause, which tends to affect less massive subhalos, 
is that halos can be ``flybys'': they can
enter the physical space associated with a more massive FoF halo and become temporarily
assigned to it as a subhalo, but later re-emerge as a separate halo.
The physical state of flybys can be complex, and their bound gas mass especially can change rapidly 
while they are in the process of interacting with the more massive halo. 

In addition to these complexities,
$M_{\mathrm{200c}}$ is not well-defined for individual satellite subhalos.
We therefore exclude from our sample at each snapshot a subset 
of the galaxies with non-central progenitors,
using the following criteria for different categories of non-central progenitors:

\begin{enumerate}

\item \textbf{A progenitor is the satellite of a central that 
is also a progenitor.} For main progenitors, 
this corresponds to the case of a central merging with 
a satellite described above.
For non-main progenitors, it indicates 
the merger of two FoF halos, and the subsequent merger
between their central and satellite subhalos.
We do not remove progenitors of this type.

\item \textbf{The main progenitor is a satellite that swaps
places with the central subhalo of its FoF halo, such that the former central is now a satellite.}
This can occur in the case of interacting galaxies/subhalos.
It can lead to a significantly undermassive main progenitor branch
when the two interacting subhalos are exchanging mass but have not
yet merged, and the initially less
massive subhalo is currently the central.
We thus entirely exclude galaxies with such main progenitors, 
reducing the sample size by $2.3\%$ (from 9543 to 9328 galaxies).

\item \textbf{A progenitor is temporarily the satellite of a central
that remains part of a distinct FoF halo.} This occurs in the case of a flyby. 
We exclude those main progenitors that are flybys at a given
timestep \textit{only at that particular timestep}.
If the main progenitor is a central subhalo at earlier or later timesteps, 
then we include the main progenitor and its $z = 0.1$ descendant in our sample 
for those timesteps. Less than $2\%$ of main progenitors are excluded as flybys at each
timestep with $z \le 4$. When we examine the combined properties of all the progenitor subhalos 
rather than only those of the main progenitor,
we perform a similar exclusion
if there exists \textit{any} progenitor at a given timestep
that is a flyby. Less than $4\%$ of 
the sample is excluded by this criterion at each $z \le 4$.

\item \textbf{A non-main progenitor is a satellite whose
central is not a progenitor but becomes a satellite of the $\mathbf{z = 0.1}$ descendant.}
This results from the merger of two FoF halos in which a 
satellite of the less massive halo has merged with the central of the more massive
halo. This satellite contributes to the central
stellar mass of the new FoF halo created by the merger, 
but its former central does not.
When considering the aggregate properties of all the progenitor subhalos, 
we exclude at each individual timestep those subhalos with progenitors
of this type. The maximum fraction of objects excluded
by this criterion is $13.2\%$, at $z = 2.24$.
Because of the substantial fraction of galaxies excluded by this criterion,
we comment on its impact on our results during their presentation in \S\hyperref[sec:results2]{3.2}.

\end{enumerate}

We also note that the FoF halo hosting the main progenitor
at each timestep may contain flyby subhalos that
are not present in the FoF halo hosting the $z = 0.1$ descendant. We do not correct for this as
we expect these subhalos to generally constitute little of
the total mass of the FoF halo, but they will contribute some scatter 
to the correlation between progenitor and descendant properties.

We match the main progenitors of the galaxies in our sample
to the corresponding subhalos in the DMO simulation, in 
the same manner as for our $z = 0.1$ sample.
We do this at a subset of redshift snapshots: $z = 0.27, 0.50, 0.74, 1.00, 1.50$ and $2.00$.
At $z = 2.00$, the main progenitor host halo masses are typically $\sim 1/4$ of the mass
of the host halos of the descendants, but with a very large scatter; 
$99.7\%$ of the main progenitors have host halo masses above $10^{10} \mathrm{M_{\odot}}$, 
which contain over 1000 particles. 
Once the cuts described above have been applied
to the main progenitor sample, 
we are able to match $96-99\%$ of the progenitors to the DMO simulation
at the selected redshifts. 
Because subhalos in the DMO simulation can also become flybys, 
we exclude DMO matches that are satellites whose centrals 
do not match to any subhalo within the corresponding FoF host halo
in the reference simulation. Less than $1\%$ of the DMO matches are
excluded by this criterion.

\begin{figure}
  \begin{center}
    \includegraphics[width=\columnwidth]{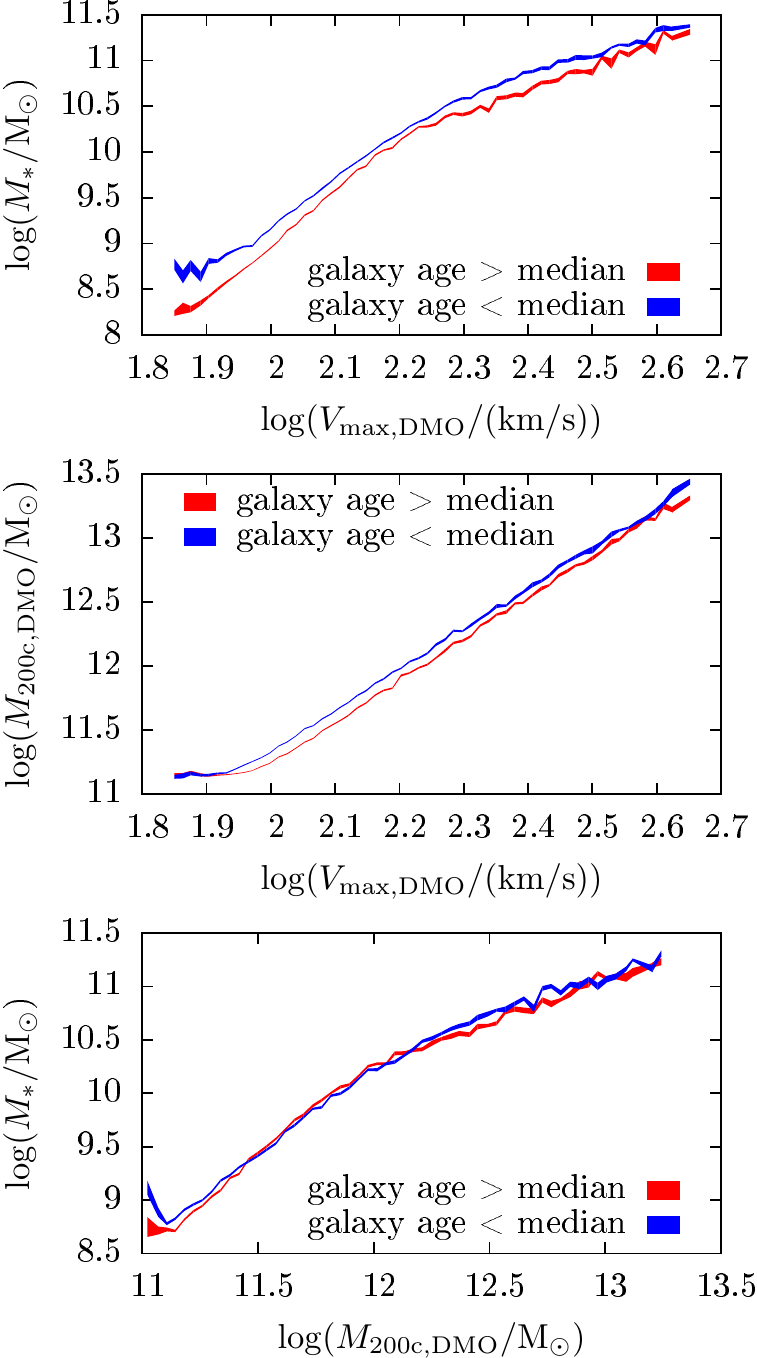}
    \caption{\textit{Top Panel:} The relationship between the stellar mass, $M_{*}$, of central galaxies,
and the maximum circular velocity of the matched dark matter halo in the dark-matter only
simulation (see text), $V_{\mathrm{max, DMO}}$. 
In each of 90 fine bins in $V_{\mathrm{max, DMO}}$, the red line
shows the mean $M_{*}$ of galaxies above the median stellar population
age in the bin, while the blue line is the same for galaxies below the median.
The thickness of the lines
represents the error on the mean $M_{*}$ in each bin.
Galaxies with older stellar population ages have lower stellar masses,
on average, at fixed $V_{\mathrm{max, DMO}}$.
\textit{Middle Panel:} Same as the top panel, but showing the DMO halo mass $M_{\mathrm{200c, DMO}}$ 
on the vertical axis rather than the central stellar
mass of the galaxy. Central galaxies with older stellar population ages
are associated with less massive (i.e., more concentrated) halos at fixed $V_{\mathrm{max, DMO}}$. 
This is a reflection of the influence of halo assembly time,
which is highly positively correlated with halo concentration, 
on the age of the central galaxy. 
\textit{Bottom Panel:} The mean central galaxy stellar mass $M_{*}$ 
as a function of the DMO halo mass, $M_{\mathrm{200c, DMO}}$, again split by the median galaxy stellar population age
in each bin. There is little correlation 
between $M_{*}$ and galaxy age at fixed halo mass.}
    \label{graph:age_split}
  \end{center}
\end{figure}

\begin{figure*}
  \begin{center}
    \includegraphics[width=\textwidth]{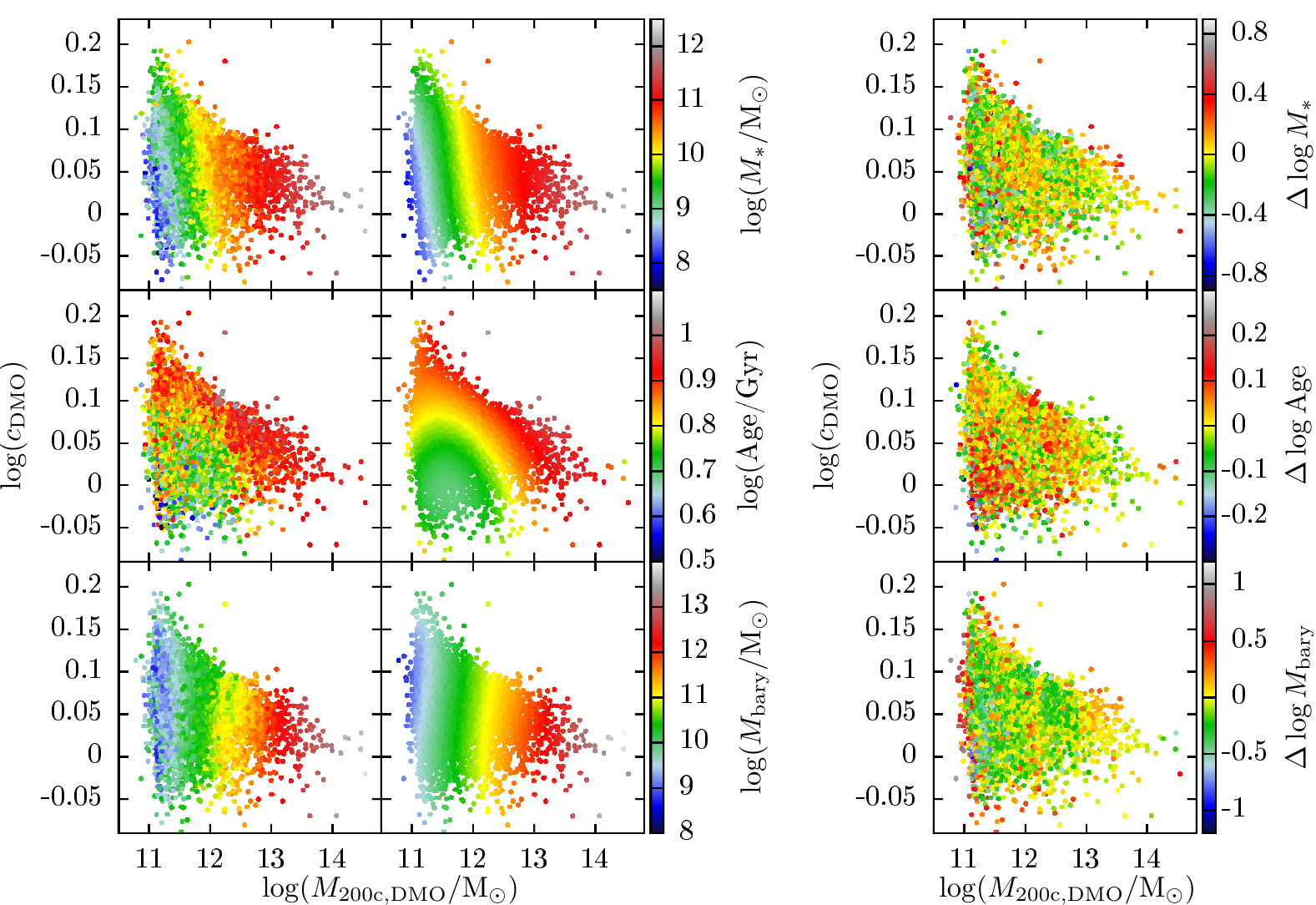}
    \caption{Galaxy/halo properties as a function of $M_{\mathrm{200c, DMO}}$ and 
$c_{\mathrm{DMO}} \equiv V_{\mathrm{max, DMO}}/V_{\mathrm{200c, DMO}}$
of the matched halo in the DMO simulation (see text).
\textit{Leftmost panels:} Points are colored by the following properties, from top to bottom: central galaxy stellar mass, 
central galaxy mean stellar population age, and total bound baryonic mass (gas plus stars) within the halo (including substructure). 
\textit{Middle panels:} Same as the leftmost panels, but now smoothed via a smoothing spline to obtain the mean relation
as a function of $M_{\mathrm{200c, DMO}}$ and $c_{\mathrm{DMO}}$.
\textit{Rightmost panels:} The difference of the leftmost and middle panels, showing the scatter in each galaxy/halo
property, denoted by ``$\Delta$'' (see also Eqn. \ref{delta_eq}).}
    \label{graph:halos_colors}
  \end{center}
\end{figure*}

Throughout the results section, we refer to dark matter halo properties from the 
DMO simulation using the subscript ``DMO'', whereas those without this subscript are
taken from the reference simulation.
$M_{\mathrm{200c}}$ refers to the mass within the radius within which the mean overdensity is 200 times the critical
density, and $M_{\mathrm{dark}}$ is used to refer to the total mass in dark matter
particles assigned to a FoF halo.

We use as a proxy for the NFW halo concentration parameter $c = R_{200}/R_{\mathrm{s}}$ 
the ratio $V_{\mathrm{max}}/V_{200}$ \citep{prada2012}. Here $V_{\mathrm{max}}$ is the 
maximum circular velocity and $V_{200} = (GM_{200}/R_{200})^{1/2}$.
We note, however, that because the maximum circular velocity of each central subhalo 
is computed by SUBFIND, it 
does not include the mass contribution of any other subhalos inside the FoF halo; as a result,
in a minority of cases ($4\%$ of our sample), $V_{\mathrm{max, DMO}}/V_{\mathrm{200c, DMO}} < 1$.

\section{Results}

\subsection{Stellar mass scatter at $z = 0.1$}
\label{sec:results1}

In \citet{matthee2017} it was found that the stellar mass, $M_{*}$, of central galaxies correlated well
with the maximum circular velocity of their matched DMO halos, $V_{\mathrm{max, DMO}}$. The authors
investigated whether the residual scatter in this relation correlated with any other DMO
halo properties, including concentration and assembly time, finding that it did not. 
Here we attempt to identify the origin of this scatter by considering correlations
with baryonic galaxy properties. We find that the 
scatter in $M_{*}$ \textit{does} correlate with the mean age of the
 stellar population of the galaxy. This can be seen in
the top panel of Figure \ref{graph:age_split}, which plots the mean stellar mass 
in fine bins of $V_{\mathrm{max, DMO}}$, split by the median galaxy
stellar population age in each bin. The thickness of the lines shows the error on the mean --- the scatter in $M_{*}$ for galaxies
above and below the median age is significant, but there is a clear offset in their mean $M_{*}$, such
that galaxies with younger stellar populations have higher stellar masses at fixed $V_{\mathrm{max, DMO}}$.

The middle panel shows the same bins in $V_{\mathrm{max, DMO}}$, again split by the median stellar age in 
each bin, but now versus the halo mass of each galaxy's matched DMO halo, $M_{\mathrm{200c, DMO}}$. 
The halo mass is related to the halo concentration at fixed
$V_{\mathrm{max, DMO}}$, such that 
less massive halos have higher concentrations (indeed, for a perfect NFW halo profile, $V_{\mathrm{max}}$
is simply an increasing function of $M_{\mathrm{200c}}$ and concentration). 
A higher halo concentration is highly correlated
with an earlier halo formation time \citep{wechsler2002},
implying that halos with lower $M_{\mathrm{200c, DMO}}$ at fixed $V_{\mathrm{max, DMO}}$ have earlier
assembly times.

In the middle panel, we see that galaxies with younger
stellar populations have more massive (less concentrated, later-forming) halos 
at fixed $V_{\mathrm{max, DMO}}$. This implies a positive correlation between halo
age and galaxy age at fixed $V_{\mathrm{max, DMO}}$, as might be expected. 
However, in \citet{matthee2017}, it was found that there is no correlation between $M_{*}$ and concentration
or halo formation time at fixed $V_{\mathrm{max, DMO}}$. Thus, the age difference seen in the middle panel 
of Figure \ref{graph:age_split} has no
correlation with the stellar mass of the galaxy, and is uncorrelated with the trend in the top panel. 

The bottom panel shows the relation between halo mass and stellar mass --- i.e. the stellar-halo
mass relation --- split by galaxy stellar population age. The trend seen here is a combination of the trends seen
 in the top two panels. At fixed $M_{\mathrm{200c, DMO}}$, halos have a range of values
of $V_{\mathrm{max, DMO}}$. Those with higher $V_{\mathrm{max, DMO}}$ have on 
average central galaxies with higher $M_{*}$; furthermore, 
 the galaxies are older on average, as seen in the middle panel.
If these were the only trends present, there would be a positive correlation
between galaxy stellar mass and stellar population age at fixed $M_{\mathrm{200c, DMO}}$. However,
 there is an additional inverse correlation between $M_{*}$ and stellar population age at fixed 
$V_{\mathrm{max, DMO}}$, as seen in the top panel. The combination of these two opposing trends
results in a lack of significant correlation between galaxy stellar mass and stellar population age
at fixed $M_{\mathrm{200c, DMO}}$.

\begin{figure}
  \begin{center}
    \includegraphics[width=\columnwidth]{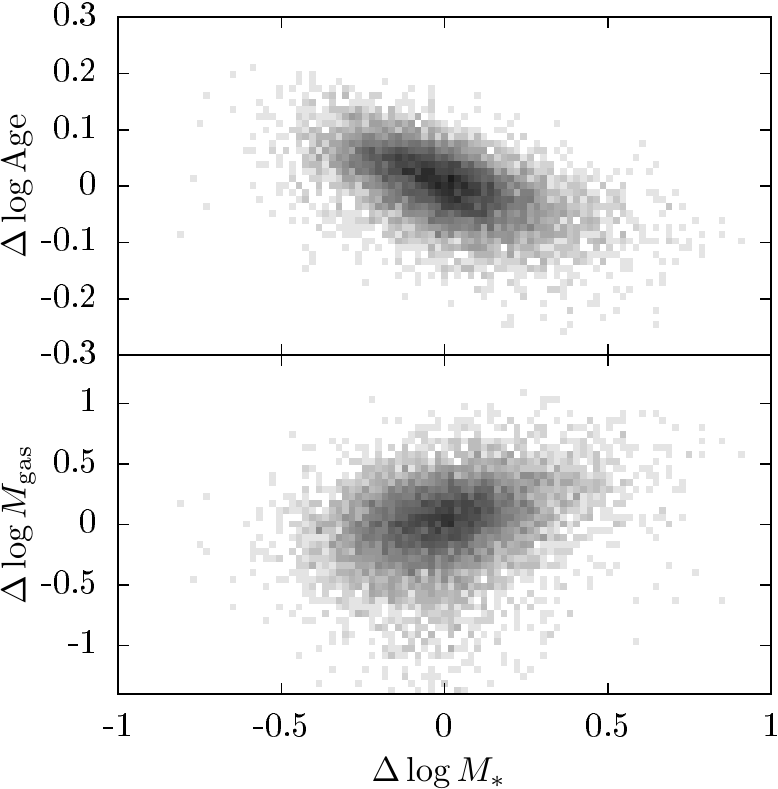}
    \caption{\textit{Top Panel:} The deviation from the mean value at fixed $M_{\mathrm{200c, DMO}}$
and $c_{\mathrm{DMO}}$ of the stellar mass ($\Delta \log M_{*}$) versus 
the deviation from the mean stellar population age ($\Delta \log \mathrm{Age}$). (See
Eqn. \ref{delta_eq} and text for details.) The darkness of the shade represents
the log-density of points in each bin.
\textit{Bottom Panel:} $\Delta \log M_{*}$ versus the deviation from the mean of the total gas
mass inside the galaxy's host halo, $\Delta \log M_{\mathrm{gas}}$.}
    \label{graph:delta_age}
  \end{center}
\end{figure}

We now understand how $M_{*}$ varies
as a function of halo mass and concentration, which are the two ``most important'' halo parameters
with which most other halo parameters are highly correlated \citep{jeeson2011,
 skibba2011, wong2012}.
Therefore, we wish to remove the mean dependence of $M_{*}$ 
and other galaxy properties on the halo mass and concentration and consider the
correlations between deviations from the mean.  The manner in which we do this is demonstrated in 
Figure \ref{graph:halos_colors}. The leftmost panels plot 
$c_{\mathrm{DMO}} \equiv V_{\mathrm{max, DMO}}/V_{\mathrm{200c, DMO}}$, 
a proxy for the halo concentration (see \S\hyperref[sec:sample]{2.2}),
versus the DMO halo mass $M_{\mathrm{200c, DMO}}$.
 Each halo is color-coded
by the value of one of its baryonic properties --- from top to bottom: central galaxy stellar mass $M_{*}$, 
central galaxy mean stellar population age, 
and total bound baryonic mass in the halo. From these plots various mean trends are evident: the stellar mass follows lines
of constant $V_{\mathrm{max, DMO}}$, $M_{\mathrm{bary}}$ correlates primarily with $M_{\mathrm{200c, DMO}}$, 
and stellar population age traces a more complex increasing function of both 
halo mass and concentration.

\begin{figure*}
  \begin{center}
    \includegraphics[width=\textwidth]{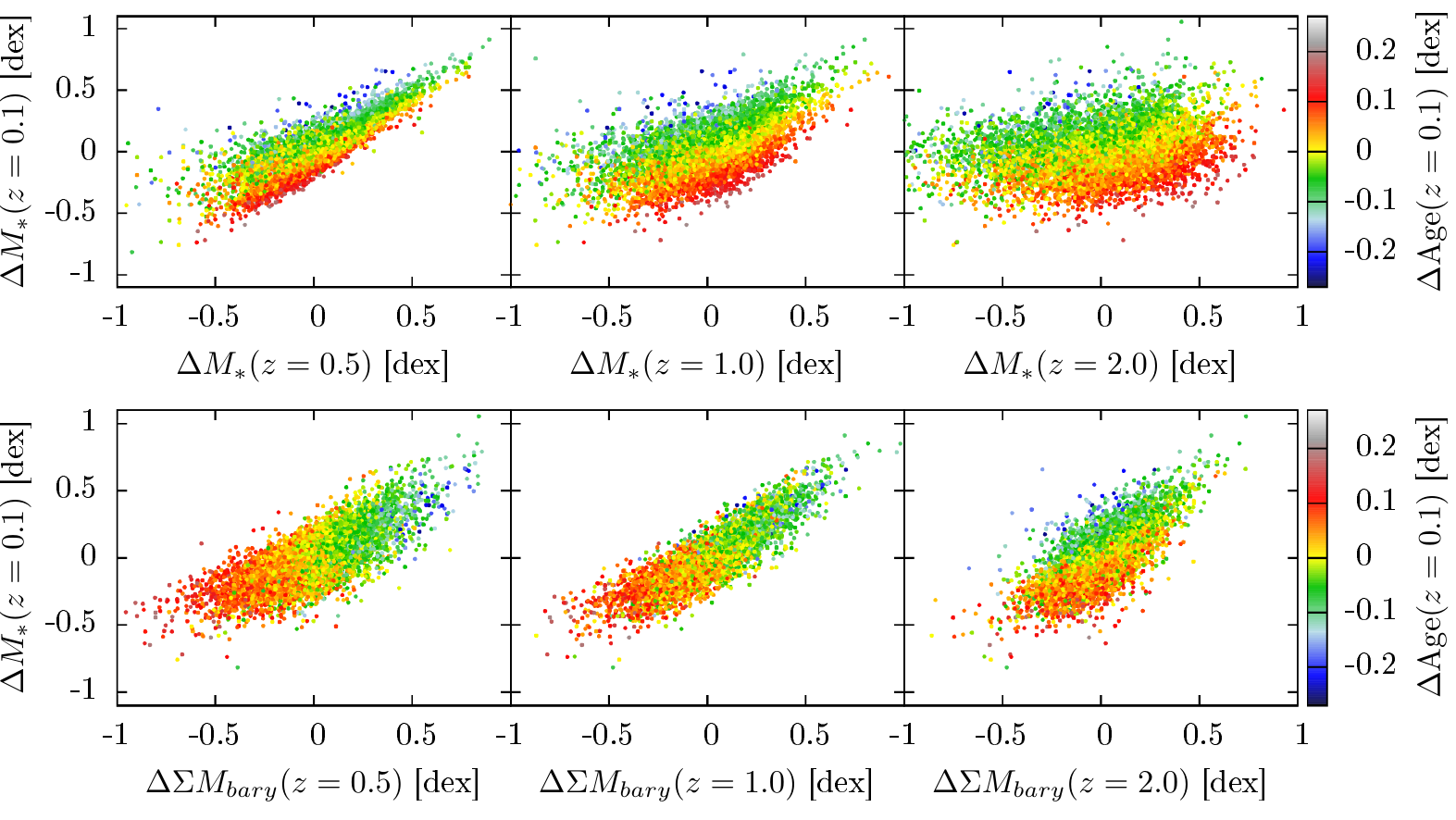}
    \caption{As in Figure \ref{graph:delta_age}, the deviation of various galaxy and halo properties from
the mean at fixed $z = 0.1$ $M_{\mathrm{200c, DMO}}$ and $c_{\mathrm{DMO}}$ 
(see Eqn. \ref{delta_eq} and text of \S\hyperref[sec:results1]{3.1} for more details). 
The top panels show $\Delta \log M_{*}$ for the $z = 0.1$ galaxy sample 
versus $\Delta \log M_{*}$ of their main progenitor
galaxies at $z = 0.5$ (left), $1.0$ (middle), and $2.0$ (right). 
Points are colored by $\Delta \log \mathrm{Age}$ at $z = 0.1$, where the 
Age refers to the stellar population age of each galaxy. The bottom panels
show the same, but for $\Delta \log M_{*} (z = 0.1)$ versus 
$\Delta \log \Sigma M_{\mathrm{bary}}$ at $z = 0.5, 1.0$, and $2.0$, where 
$\Sigma M_{\mathrm{bary}}$ is the sum of the stellar and gas masses of all the progenitors
of each galaxy. For $z \gtrsim 1$, 
$\Sigma M_{\mathrm{bary}}$ of the progenitor halos is a better 
predictor of $\Delta \log M_{*} (z = 0.1)$ than $\Delta \log M_{*}$ of the main progenitor galaxy.}
    \label{graph:deltas}
  \end{center}
\end{figure*}

We compute the mean dependence of each parameter on $M_{\mathrm{200c, DMO}}$ and $c_{\mathrm{DMO}}$
by fitting a bivariate smoothing spline in log-space.  We do not find that varying the smoothing parameters 
has a large effect on our results, and simply subtracting a mean in bins of 
$\log(M_{\mathrm{200c, DMO}})$ and $\log(c_{\mathrm{DMO}})$ produces 
consistent results. These mean relations are shown in the middle set of panels in Figure \ref{graph:halos_colors}.
We then define the deviation from this mean for $M_{*}$ as 
\begin{equation}
\label{delta_eq}
\Delta \log M_{*} \equiv \log(M_{*}) - \overline{\log(M_{*})} (\log(M_{\mathrm{200c, DMO}}), \log(c_{\mathrm{DMO}}))
\end{equation}
and similarly for the other galaxy/halo parameters.
The deviations from the mean produced by subtracting the 
middle panels from the leftmost panels of Figure \ref{graph:halos_colors}
is shown in the rightmost panels.

In Figure \ref{graph:delta_age}, we plot the deviation of the 
central galaxy stellar population age from the mean relation, $\Delta \log \mathrm{Age}$, versus $\Delta \log M_{*}$,
confirming that there is a negative correlation (Spearman correlation coefficient $R_{s} = -0.55$)
between the two as could be inferred from Figure \ref{graph:age_split}.
In the bottom panel of Figure \ref{graph:delta_age}, we plot $\Delta \log M_{*}$ versus $\Delta \log M_{\mathrm{gas}}$,
where the latter is computed using all the gas in the FoF host halo
that is bound to any substructure. There is a weak positive correlation ($R_{s} = 0.29$)
between $\Delta \log M_{*}$ and $\Delta \log M_{\mathrm{gas}}$,
such that halos whose central galaxies have above-average stellar masses also tend to have a slight excess of gas relative to
similar halos. Interestingly, this implies that such halos tend to contain a higher
overall baryonic mass relative to other halos of the same mass and concentration.
 
\subsection{Correlation of stellar mass scatter with progenitor properties}
\label{sec:results2}

To understand the origin of the scatter in the $V_{\mathrm{max}}-M_{*}$ relation at $z = 0.1$, 
we attempt to correlate the scatter
 to the properties of the progenitors of the galaxies. 
The selection of the progenitors
and the cuts made to our sample are described in \S\hyperref[sec:sample]{2.2}. As for the stellar population age
and gas mass above, we examine differences in the progenitor properties after removing
the mean dependence on the halo mass and concentration of the descendant $z=0.1$ halos,
denoting this with a ``$\Delta$'' in front of the property. In this way we examine
the variation in the growth histories of galaxies and halos with the same present-day
properties and how this affects the stellar mass of their central galaxies.

We consider the properties of the main progenitor branch (defined in \S\hyperref[sec:sample]{2.2}), 
including the stellar mass of the main progenitor
galaxy, the total baryonic mass\footnote{We compute the baryonic mass of the 
main progenitor host as the sum of the gas
and stellar masses bound to each subhalo in the FoF halo that hosts the main progenitor galaxy.
However, this halo may contain subhalos that do not merge with the central galaxy by $z = 0.1$ and
are thus not its progenitors. In practice,
this is a minor difference because the gas of satellite subhalos is generally stripped
quickly upon entering a FoF halo and is reassigned to the central subhalo, and also because the
satellite galaxies that take a long time to merge with the central tend to have low masses.} 
within the halo hosting said galaxy, and 
the halo mass of the corresponding DMO halo.
We also look at the sum of the stellar, baryonic, and dark matter masses
of all the progenitor subhalos of each $z = 0.1$ galaxy/subhalo
at different redshifts.

\begin{figure}
  \begin{center}
    \includegraphics[width=\columnwidth]{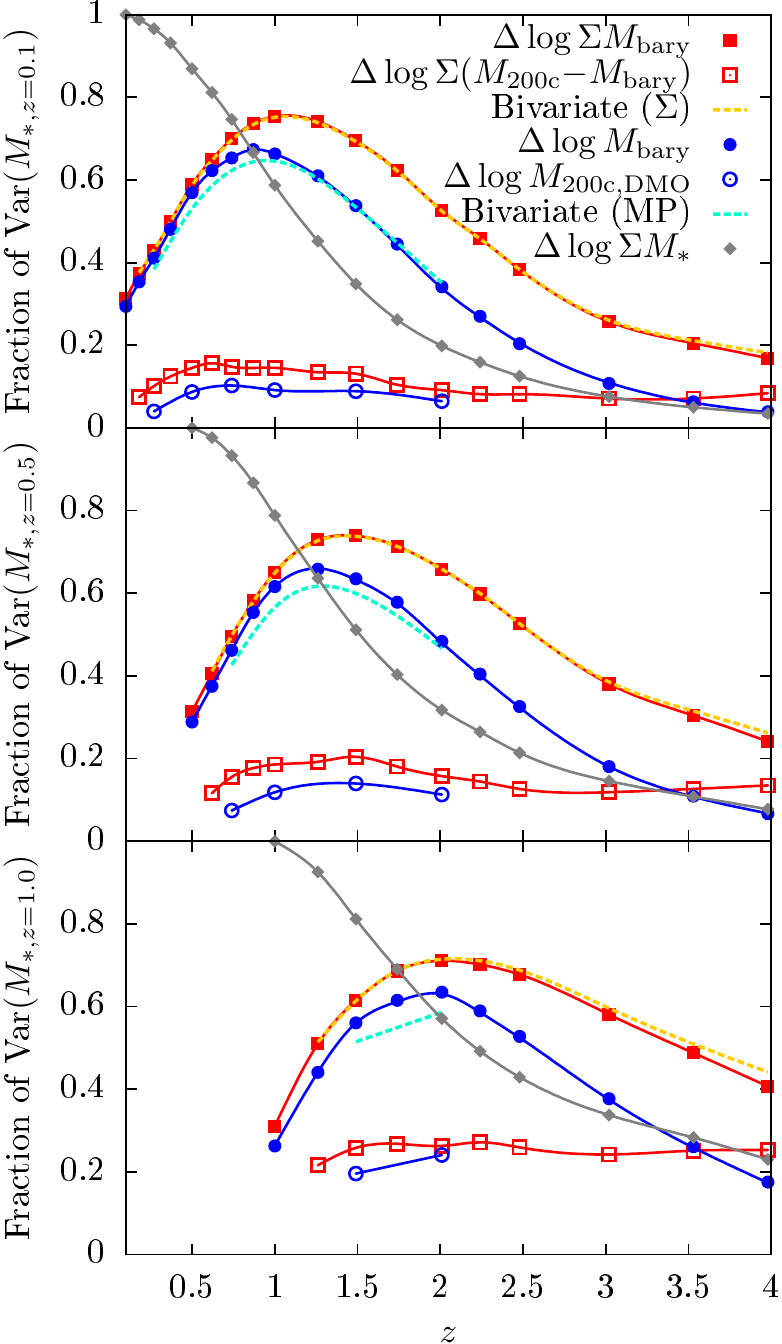}
    \caption{The fraction of the variance of $\Delta \log M_{*}$ for central galaxies
 with $M_{\mathrm{200c}} > 10^{11} \mathrm{M_{\odot}}$ at $z = 0.1$ (top), 
$z = 0.5$ (middle), and $z = 1.0$ (bottom) that can be
accounted for by the scatter in various properties of their progenitors as
a function of progenitor redshift.
(See Eqn. \ref{delta_eq} and \S\hyperref[sec:results1]{3.1}
for an explanation of the notation.) 
The gray curve with diamond points
corresponds to $\Delta \log \Sigma M_{*}$, where $\Sigma M_{*}$ is the sum of the stellar masses of 
all the progenitors of each galaxy. The red line with filled square points shows 
$\Delta \log \Sigma M_{\mathrm{bary}}$, where $\Sigma M_{\mathrm{bary}}$ is the sum of the baryonic masses
 ($M_{\mathrm{gas}} + M_{*}$) of the progenitors at each redshift.
The red line with open square points corresponds to $\Delta \log \Sigma (M_{\mathrm{200c}} - M_{\mathrm{bary}})$, where
$\Sigma (M_{\mathrm{200c}} - M_{\mathrm{bary}})$ is the sum of the total halo masses of each
galaxy's progenitors, minus the mass of their baryonic components.
The orange dashed line shows the contribution to the variance from the linear combination of
$\Delta \log \Sigma (M_{\mathrm{200c}} - M_{\mathrm{bary}})$ and $\Delta \log f_{\mathrm{bary}, \Sigma}$, 
where $f_{\mathrm{bary}, \Sigma} \equiv \Sigma M_{\mathrm{bary}}/\Sigma M_{\mathrm{200c}}$.
The blue line with filled circular points shows $\Delta \log M_{\mathrm{bary}}$, where $M_{\mathrm{bary}}$
is the baryonic mass within the host halo of the main progenitor galaxy.
The blue curve with open circular points corresponds to 
$\Delta \log M_{\mathrm{200c, DMO}}$, where $M_{\mathrm{200c, DMO}}$ is the mass of the 
DMO halo corresponding to the host halo of the main progenitor galaxy in
the reference simulation. The cyan dashed line represents the combined contribution of 
$\Delta \log M_{\mathrm{200c, DMO}}$ and $\Delta \log f_{\mathrm{bary, MP}}$, where
$f_{\mathrm{bary, MP}} \equiv M_{\mathrm{bary}}/M_{\mathrm{200c}}$.}
    \label{graph:redshift_correlation}
  \end{center}
\end{figure}

We denote the baryonic mass of the FoF halo hosting the main progenitor galaxy as $M_{\mathrm{bary}}$ 
and the sum of the baryonic masses of all the progenitor subhalos as $\Sigma M_{\mathrm{bary}}$. 
Similarly, $M_{*}$ refers to the stellar mass of the main progenitor and $\Sigma M_{*}$ to the
sum of the stellar masses of all the progenitors. We match the main progenitor subhalos
at selected redshifts to the
corresponding subhalos in the DMO simulation, as described in \S\hyperref[sec:sample]{2.2},
and refer to the mass of the host FoF halo as $M_{\mathrm{200c, DMO}}$.
We do not attempt to match the full sample of all progenitor
subhalos because many are low-mass and it is more difficult to obtain accurate matches between
the two simulations for low-mass subhalos. We do utilize the sum of the $M_{\mathrm{200c}}$ halo 
masses from the reference simulation, minus the baryonic component, 
denoting this as $\Sigma (M_{\mathrm{200c}} - M_{\mathrm{bary}})$.

In the top row of panels in Figure \ref{graph:deltas}, we show a comparison of 
$\Delta \log M_{*}$ at $z = 0.1$ to $\Delta \log M_{*}$ of the main progenitor galaxy 
at $z = 0.5, 1.0$, and $2.0$ (all computed relative to $M_{\mathrm{200c, DMO}}$
and $c_{\mathrm{DMO}}$ of the descendant at $z = 0.1$). 
Unsurprisingly, those galaxies with atypically high stellar masses at $z = 0.1$ tend to also have progenitors with high
stellar masses. The correlation decreases with increasing redshift:
the Spearman correlation coefficient is
$R_{s} = 0.85$ at $z = 0.5$, $0.59$ at $z = 1.0$, and $0.25$ at $z = 2.0$. 
The points are color-coded
by $\Delta \log \mathrm{Age}$ at $z = 0.1$, which follows a diagonal trend in the top panels
because it is correlated with the mass of stars formed between the redshift of that 
panel and $z = 0.1$.

It is interesting to compare the top panels of Figure \ref{graph:deltas} to the bottom ones, which show
$\Delta \log \Sigma M_{\mathrm{bary}}$ computed for the same redshifts as the top panels.
Here we see that $\Delta \log M_{*}$ at $ z = 0.1$ is positively correlated with $\Delta \log M_{\mathrm{bary}}$
at each redshift,
with $R_{s} = 0.75$ at $z = 0.5$, $0.86$ at $z = 1.0$, and $0.69$ at $z = 2.0$.
Unlike for the stellar mass in the top panels,
the correlation strengthens between $z = 0.5$ and $z = 1.0$,
and for $z \gtrsim 1$ the correlation between $\Delta \log M_{*} (z = 0.1)$ and $\Delta \log \Sigma M_{\mathrm{bary}}$
is stronger than than that between
$\Delta \log M_{*} (z = 0.1)$ and $\Delta \log M_{*}$. Although the stellar
mass of the progenitors is part of $M_{\mathrm{bary}}$, the correlation
between $\Delta \log M_{*} (z = 0.1)$ and $\Delta \log \Sigma M_{\mathrm{bary}}$ at higher redshifts
is mainly driven by the gas mass, as will be shown below.

In the bottom panels of Figure \ref{graph:deltas}, it is also apparent that
for $z \lesssim 1$, the mean stellar age of the galaxy at $z = 0.1$ is negatively
correlated with $\Delta \log \Sigma M_{\mathrm{bary}}$. This reveals the origin
of the negative correlation between stellar mass and mean stellar population age at $z = 0.1$.
It is possible for two sets of halo progenitors at $z \sim 1$ with different total baryonic masses to 
evolve into halos with the same $M_{\mathrm{200c, DMO}}$
and $c_{\mathrm{DMO}}$ at $z = 0.1$; however, due to their different
initial baryonic masses, they will experience different amounts of star formation at $z < 1$
and the one with higher initial baryonic mass will tend to have a younger, more massive central galaxy. 

The relationship between $\Delta \log M_{*} (z = 0.1)$ and progenitor
properties is revealed in greater detail in Figure \ref{graph:redshift_correlation}.
In the top panel, we show the fraction of the variance of $\Delta \log M_{*}$ at $z = 0.1$
that can be accounted for by  
different progenitor properties as a function of redshift. 
This is done by fitting a line to the relationship between each progenitor property
and $\Delta \log M_{*} (z=0.1)$, defined by $f(x) = ax$ (the intercept is taken
to be zero because all properties are normalized by removing the mean at fixed
$M_{\mathrm{200c, DMO}}$ and $c_{\mathrm{DMO}}$).
The fractional contribution to the variance is 
 $[ \mathrm{Var}(\Delta \log M_{*, z=0.1}) - \mathrm{Var} (\Delta \log M_{*, z=0.1} - ax ) ] / \mathrm{Var}(\Delta \log M_{*, z=0.1})$.
Here $\mathrm{Var}(\Delta \log M_{*,z=0.1})$ varies slightly for the different redshift points
due to the different sample cuts
at each point (see \S\hyperref[sec:sample]{2.2}) but is always $\approx (0.185$ dex$)^{2}$.

The red line with filled square points in the
top panel of Figure \ref{graph:redshift_correlation} shows
the fraction of the variance of $\Delta \log M_{*} (z=0.1)$ accounted for by $\Delta \log \Sigma M_{\mathrm{bary}}$
at each redshift --- the quantity that was plotted 
along the x-axis in the bottom panels of Figure \ref{graph:deltas}. The correlation between $\Delta \log M_{*} (z=0.1)$
and $\Delta \log \Sigma M_{\mathrm{bary}}$ peaks at $z \approx 1.1$, where the baryonic
mass of the progenitors accounts for $75\%$ of the variance of $\Delta \log M_{*}$ at $z = 0.1$. For comparison,
we show as the gray line with diamond points $\Delta \log \Sigma M_{*}$, where 
$\Sigma M_{*}$ is the sum of the stellar masses of the progenitor galaxies at each
redshift. (Note that this is different from what is plotted in the top panels of Figure \ref{graph:deltas}, which
shows only the stellar mass of the main progenitor galaxy). 
For $z \gtrsim 0.8$, the total baryonic mass accounts for a
larger fraction of the scatter in $M_{*}$ at $z=0.1$ than $\Sigma M_{*}$.
This indicates that the gas reservoir available for star formation at early times is the major factor
determining the eventual stellar mass of the central galaxy in a halo.

The blue line with filled circular points
is the same as the red line with filled square points, but includes only the baryonic content of the host halo of the 
main progenitor galaxy. The baryon content within the host halo
of the main progenitor galaxy (which is also generally the most massive progenitor halo)
accounts for $67\%$ of the variance of $\Delta \log M_{*} (z = 0.1)$ at $z \approx 0.9$, meaning that
the properties of the main progenitor halo alone account for the majority ($89\%$) of the variance that is accounted
for by all the progenitors. However, due to our
chosen lower halo mass cut of $10^{11} \mathrm{M_{\odot}}$ and the steepness of the 
halo mass function, the typical halo in our sample has fairly low mass and
consequently does not gain a significant fraction of its mass from mergers.
We present results for different halo masses later in this section.

The scatter in $\Delta \log M_{\mathrm{bary}}$
is partly due to the scatter in progenitor halo masses, 
since higher-mass halos have, on average, higher baryonic masses. 
We plot in the top panel of Figure \ref{graph:redshift_correlation}, 
as the blue line with open circular points,
the contribution to the variance of $\Delta \log M_{*} (z = 0.1)$
by scatter in the main progenitor DMO halo mass,
$M_{\mathrm{200c, DMO}}$. We see that the variance in the progenitor halo mass alone
is only able to account for a small fraction ($\lesssim 10\%$) of the variance in $\Delta \log M_{*} (z=0.1)$.
Similarly, the red line with open square points shows the contribution to the variance in $M_{*}$ 
by the variance in the sum of the $M_{\mathrm{200c}}$ progenitor halo masses
from the reference simulation, minus their baryonic component, denoted $\Sigma (M_{\mathrm{200c}} - M_{\mathrm{bary}})$.
Again, the variance in the dark matter mass of the progenitors can only account for
$\approx 15\%$ of the variance in $\Delta \log M_{*} (z = 0.1)$.

The remainder of the variance in the progenitor baryonic masses
can be thought of as resulting from variation in the baryon mass fraction of
halos, $M_{\mathrm{bary}}/M_{\mathrm{200c}}$. For the main
progenitor host halo, $f_{\mathrm{bary, MP}} \equiv M_{\mathrm{bary}}/M_{\mathrm{200c}}$,
and for the aggregate of all the progenitors,
$f_{\mathrm{bary}, \Sigma} \equiv \Sigma M_{\mathrm{bary}}/\Sigma M_{\mathrm{200c}}$
is the mass-weighted average baryon mass fraction. The
average baryon mass fraction within halos is a function of halo mass, so
the scatter in the baryon fraction of progenitor halos at fixed descendant
halo properties is correlated with the scatter in the halo mass(es) of the progenitor halo(s).
Therefore, to show the additional contribution of
$\Delta \log f_{\mathrm{bary}, \Sigma}$ at fixed $\Delta \log \Sigma (M_{\mathrm{200c}} - M_{\mathrm{bary}})$,
we fit the linear combination of these two parameters ($f(x, y) = ax + by$)
to $\Delta \log M_{*} (z = 0.1)$ and plot the fraction of the variance accounted for
as the orange dashed line in the top panel of Figure \ref{graph:redshift_correlation}.

The addition of $\Delta \log f_{\mathrm{bary}, \Sigma}$ at fixed
$\Delta \log \Sigma (M_{\mathrm{200c}} - M_{\mathrm{bary}})$ is able to 
account for a for as much of the variance in $z = 0.1$ 
stellar mass as $\Delta \log \Sigma M_{\mathrm{bary}}$, but not significantly more.
This implies that the baryonic mass of the progenitor halos determines the central
stellar mass of the descendant halo, and the 
scatter in the progenitor halo mass and baryon fraction are only important
to the extent that they predict the baryonic mass.
It also confirms that the scatter in the baryonic mass within 
progenitor halos is primarily dependent on the scatter in
baryonic mass fraction at fixed progenitor halo mass.

Similarly, we plot the combined contribution
of $\Delta \log M_{\mathrm{200c, DMO}}$ and $\Delta \log f_{\mathrm{bary, MP}}$
as the dashed teal line in the top panel of Figure \ref{graph:redshift_correlation}. 
In this case the contribution to the variance in $\Delta \log M_{*} (z = 0.1)$
is slightly less than that of $\Delta \log M_{\mathrm{bary}}$ because the dark matter
mass is from the matched halo in the DMO simulation rather than the reference simulation,
and is a poorer predictor of $\Delta \log M_{\mathrm{bary}}$.
 
\begin{figure}
  \begin{center}
    \includegraphics[width=\columnwidth]{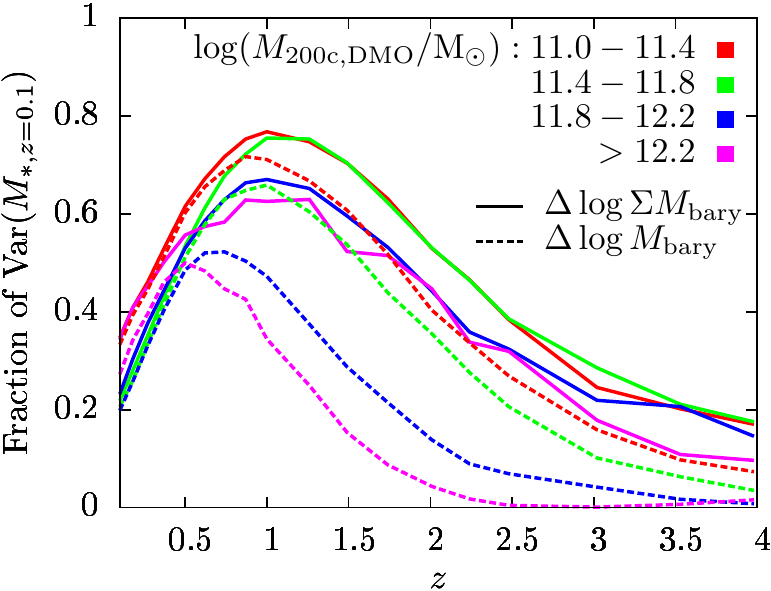}
    \caption{As in the top panel of Figure \ref{graph:redshift_correlation}, the fraction of the
variance in $\Delta \log M_{*}$ at $z = 0.1$ accounted for by $\Delta \log \Sigma M_{\mathrm{bary}}$
and $\Delta \log M_{\mathrm{bary}}$ of the progenitor galaxies, but now divided into bins of 
$z = 0.1$ $M_{\mathrm{200c, DMO}}$. Solid lines represent the contribution from $\Delta \log \Sigma M_{\mathrm{bary}}$
and dashed lines that from $\Delta \log M_{\mathrm{bary}}$. Red lines represent halos with
$M_{\mathrm{200c, DMO}}$ between $10^{11.0}$ and $10^{11.4} \mathrm{M_{\odot}}$, green lines
those between $10^{11.4}$ and $10^{11.8} \mathrm{M_{\odot}}$, blue lines those 
between $10^{11.8}$ and $10^{12.2} \mathrm{M_{\odot}}$,
and magenta lines those more massive than $10^{12.2} \mathrm{M_{\odot}}$.}
    \label{graph:mass_bins}
  \end{center}
\end{figure}
 
To check whether the correlation between $\Delta \log M_{*}$ and $\Delta \log M_{\mathrm{bary}}$ is 
specific to low redshifts, we recreate
the top panel of Figure \ref{graph:redshift_correlation}
for samples of central galaxies at $z = 0.5$ and $z = 1.0$ and their progenitors.
Specifically, we select all central galaxies at these two redshifts whose host
halos have $M_{\mathrm{200c}} > 10^{11} \mathrm{M_{\odot}}$, and match the host halos to the
corresponding halos in the DMO simulation. We perform the
same cuts to the sample described in \S\hyperref[sec:sample]{2.2}. This results in samples of 9935 and
10229 galaxies for $z = 0.5$ and $1.0$, respectively. We then recompute 
all the properties shown in the top panel of Figure \ref{graph:redshift_correlation} relative
to $M_{\mathrm{200c, DMO}}$ and $c_{\mathrm{DMO}}$
of the $z = 0.5$ and $z = 1.0$ samples. For both samples 
the variance of $\Delta \log M_{*}$ is $\approx (0.18$ dex$)^{2}$
for the full sample.

The results are shown in the lower two panels of Figure \ref{graph:redshift_correlation},
using the same symbols as in the top panel. The similarity
between the trends in the two figures implies that most of the scatter in $M_{*}$ is
produced by scatter in the baryonic masses of progenitor halos at all redshifts
up to at least $z = 1$. The redshift of peak correlation
between $\Delta \log M_{*}$ and $\Delta \log M_{\mathrm{bary}}$ or $\Delta \log \Sigma M_{\mathrm{bary}}$
is shifted by approximately the redshift difference
between the samples of galaxies. The fraction of 
the variance of $\Delta \log M_{*}$ accounted for 
by $\Delta \log \Sigma M_{\mathrm{bary}}$ at the peaks of the curves is $74\%$ for the sample of galaxies
at $z = 0.5$ and $71\%$ for that at $z = 1.0$. For $\Delta \log M_{\mathrm{bary}}$, the percentages
are $66\%$ and $63\%$. The contribution to the variance of $\Delta \log M_{*}$ from 
$\Delta \log M_{\mathrm{200c, DMO}}$ and $\Delta \log \Sigma (M_{\mathrm{200c}} - M_{\mathrm{bary}})$ appears to
be larger for higher-redshift galaxy samples, reaching $\approx 25\%$ for
the $z = 1.0$ sample, although it is still unable to 
account for the majority of the scatter.

As noted previously, our sample is dominated by low-mass halos. 
In Figure \ref{graph:mass_bins}, 
we divide our sample into four different bins of 
$z = 0.1$ $M_{\mathrm{200c, DMO}}$, and plot the contribution
to the variance in $\Delta \log M_{*} (z = 0.1)$ from 
$\Delta \log M_{\mathrm{bary}}$ of the main progenitor halo 
(dashed lines) and $\Delta \log \Sigma M_{\mathrm{bary}}$ of all the progenitors
(solid lines).
The peak contribution from $\Delta \log \Sigma M_{\mathrm{bary}}$
 decreases with increasing halo mass, 
accounting for $77\%$ of the variance in $\Delta \log M_{*} (z = 0.1)$
for halos with $10^{11.0} \mathrm{M_{\odot}} < M_{\mathrm{200c, DMO}} < 10^{11.4} \mathrm{M_{\odot}}$,
$75\%$ of the variance for halos with $10^{11.4} \mathrm{M_{\odot}} < M_{\mathrm{200c, DMO}} < 10^{11.8} \mathrm{M_{\odot}}$,
$67\%$ for $10^{11.8} \mathrm{M_{\odot}} < M_{\mathrm{200c, DMO}} < 10^{12.2} \mathrm{M_{\odot}}$, 
and $63\%$ for $M_{\mathrm{200c, DMO}} > 10^{12.2} \mathrm{M_{\odot}}$.
Interestingly, the redshift of peak correlation is $z \approx 1$ for all four
mass ranges, likely due to the fact that higher-mass halos are assembled from multiple lower-mass halos.

On the other hand, the redshift of peak correlation between $\Delta \log M_{*} (z = 0.1)$
and $\Delta \log M_{\mathrm{bary}}$ of the main progenitor halo
does vary with the halo mass range, owing to the later assembly 
time for higher-mass halos. 
The peak contribution to 
$\Delta \log M_{*} (z = 0.1)$ by $\Delta \log M_{\mathrm{bary}}$ 
also decreases significantly with halo mass, 
because higher-mass halos gain a larger fraction
of their mass from mergers with non-main progenitors.
For halos with $10^{11.0} \mathrm{M_{\odot}} < M_{\mathrm{200c, DMO}} < 10^{11.4} \mathrm{M_{\odot}}$, 
the redshift of peak correlation is $z \approx 0.9$, and
the variance in $\Delta \log M_{*} (z = 0.1)$ accounted for at this redshift is $72\%$.
Halos with $10^{11.4} \mathrm{M_{\odot}} < M_{\mathrm{200c, DMO}} < 10^{11.8} \mathrm{M_{\odot}}$
have the same redshift of peak correlation and $\Delta \log M_{\mathrm{bary}}$
accounts for a maximum of $66\%$ of the
$M_{*}$ variance. For $10^{11.8} \mathrm{M_{\odot}} < M_{\mathrm{200c, DMO}} < 10^{12.2} \mathrm{M_{\odot}}$,
these values are $z \approx 0.7$ and $52\%$, and
for $M_{\mathrm{200c, DMO}} > 10^{12.2} \mathrm{M_{\odot}}$, they are $z \approx 0.5$ and $50\%$.

In \S\hyperref[sec:sample]{2.2}, we discussed the cuts made to our sample. For the main progenitors
of our central galaxy samples (at $z = 0.1$, $0.5$, and $1.0$), the cuts remove
less than $4\%$ of the sample at each progenitor redshift, and
exclude outliers resulting from interacting galaxies and ``flyby'' progenitors.
Applying similar cuts to all the progenitors of each sample
excludes less than $7\%$ of the sample at each redshift. 
However, when considering all the progenitors,
we apply an additional cut, which excludes, at each redshift, galaxies
having a satellite progenitor whose associated central becomes
a satellite of the descendant galaxy but does not merge with it.
In these cases, $M_{\mathrm{200c}}$ 
of the FoF halo containing the satellite
progenitor is unlikely to correlate with the stellar mass of the descendant galaxy, and
the gas mass assigned to the satellite progenitor may be affected
by residing in a larger halo. 
This criterion removes a fraction of the sample as large as $13\%$ for the $z = 0.1$ sample,
$16\%$ for the $z = 0.5$ sample, and $22\%$ for the $z = 1.0$ sample.
Despite the large fraction of objects removed, this cut does not
significantly affect our results. It alters the curves in Figure \ref{graph:redshift_correlation}
by less than $4\%$ at any redshift.
This cut has a slightly larger effect at higher halo masses,
because the number of mergers and therefore the number
of satellite progenitors increases with halo mass. The peak
contribution to $\Delta \log M_{*} (z = 0.1)$ by $\Delta \log \Sigma M_{\mathrm{bary}}$
in the highest-mass bin in Figure \ref{graph:mass_bins} decreases
from $63\%$ to $56\%$ without this cut on the sample.

\begin{figure}
  \begin{center}
    \includegraphics[width=\columnwidth]{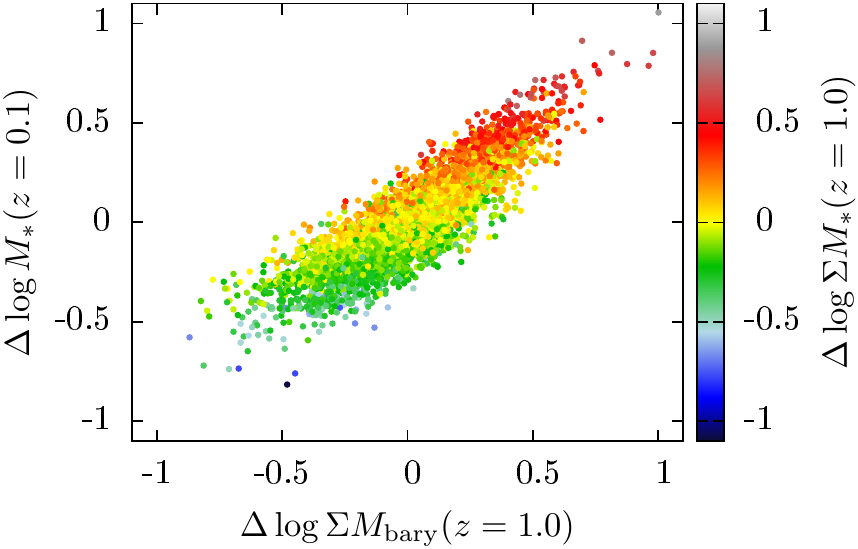}
    \caption{Same as the bottom centre panel of Figure \ref{graph:deltas}, except
now color-coded by $\Delta \log \Sigma M_{*} (z = 1.0)$, where $\Sigma M_{*}$ is the sum of the stellar 
masses of each galaxy's progenitors at $z = 1.0$. In addition to the positive correlation between
descendant stellar mass and progenitor baryonic mass at fixed $M_{\mathrm{200c, DMO}}$ and 
$c_{\mathrm{DMO}}$, those progenitors with a higher ratio of stars
to gas have descendants with higher stellar masses.} 
    \label{graph:z1_baryon_stars}
  \end{center}
\end{figure}

As shown above, scatter in the baryonic mass of progenitors produces 
most of the scatter in the $z = 0.1$ $M_{*}-V_{\mathrm{max}}$ relation. However, the 
stellar mass of the $z = 0.1$ descendants also depends somewhat on $\Delta \log M_{*}$ of the progenitors independently
of its correlation with the baryonic mass. 
Figure \ref{graph:z1_baryon_stars} shows
$\Delta \log M_{*} (z = 0.1)$ versus 
$\Delta \log \Sigma M_{\mathrm{bary}} (z = 1.0)$, colored by $\Delta \log \Sigma M_{*} (z = 1.0)$. 
The progenitor stellar and gas masses at $z = 1$ together account for a total of
$86\%$ of the variance of $\Delta \log M_{*}$ at $z = 0.1$. 
For the galaxy samples at $z = 0.5$ and $z = 1.0$, the variance accounted for
at the redshifts of peak correlation with $\Delta \log \Sigma M_{\mathrm{bary}}$
are $84\%$ and $85\%$, respectively.

\begin{figure}
  \begin{center}
    \includegraphics[width=\columnwidth]{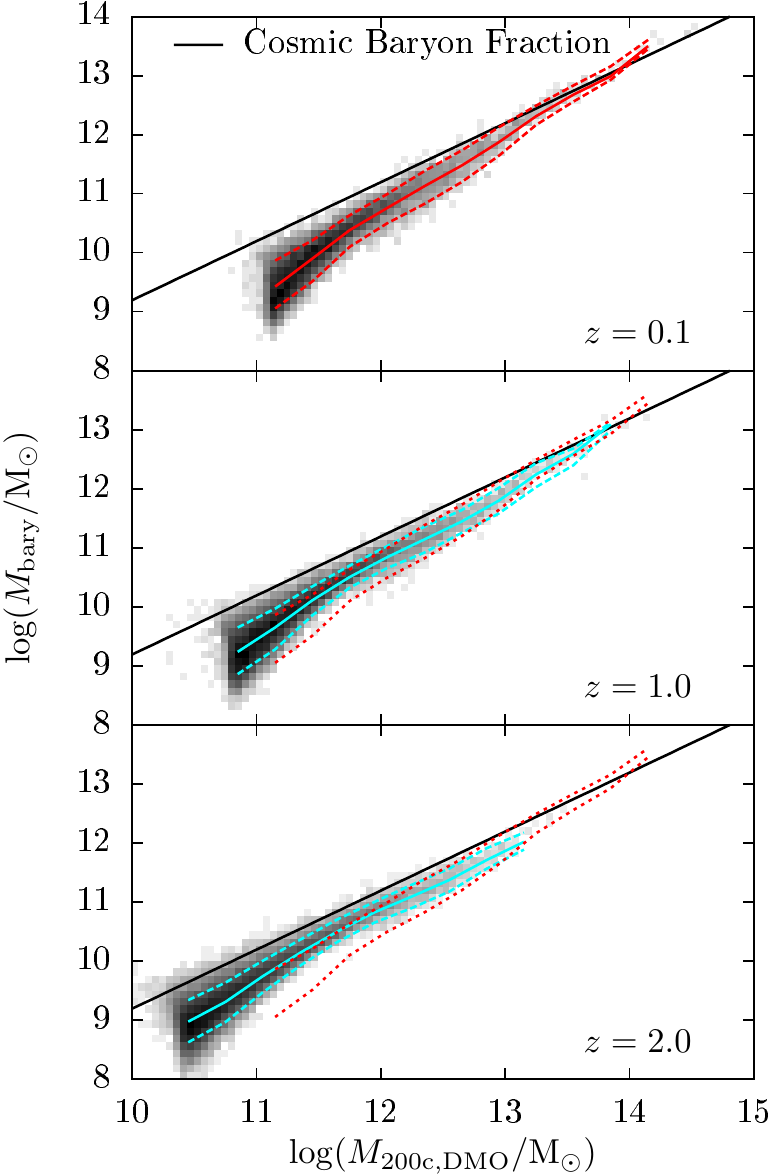}
    \caption{The total baryonic mass in each halo, including substructure, versus the matched DMO halo mass $M_{\mathrm{200c, DMO}}$. 
The different panels show this relationship at three redshifts: top $z = 0.1$, middle $z = 1.0$, and bottom
$z = 2.0$. The halo mass limits for the bottom two panels are chosen 
to approximately encompass the masses of the main progenitors
 of the halos in the top panel. The shading represents the log-density of halos in each bin. The solid black
line shows the baryonic mass expected if all halos contained the cosmic fraction of baryons. In the 
top panel, the red solid line represents the median $M_{\mathrm{bary}}$ as a function of $M_{\mathrm{200c, DMO}}$, while the 
dashed lines demarcate the bottom and top deciles. In the lower two panels, the median and deciles are represented by 
cyan lines, and the deciles from the top panel are reproduced
with red dotted lines for comparison. We are able to see that, 
for halos with $M_{\mathrm{200c, DMO}} \lesssim 10^{13} \mathrm{M_{\odot}}$,
the mean $M_{\mathrm{bary}}$ at fixed values of  $M_{\mathrm{200c, DMO}}$ decreases with time, and its scatter increases.}
    \label{graph:baryonic_mass}
  \end{center}
\end{figure}

It is important to note that all the correlations described above are
calculated at only a single redshift of the simulation. 
Since gas physics is continuous in time, 
one would expect the baryonic mass in different 
snapshots to make independent contributions to the variance in $M_{*}$.
For example, in Figure \ref{graph:z1_baryon_stars} we showed
that there is an independent correlation between $\Delta \log \Sigma M_{*} (z=1.0)$
and $\Delta \log M_{*} (z = 0.1)$ at fixed $\Delta \log \Sigma M_{\mathrm{bary}} (z = 1.0)$;
however, $\Delta \log \Sigma M_{*} (z=1.0)$ is itself highly correlated with the baryonic
masses of the galaxies' progenitors at $z \approx 2.0$, as shown in the lowest panel of Figure 
\ref{graph:redshift_correlation}.
Thus the scatter of the $z=0.1$ stellar mass in 
EAGLE can be almost entirely accounted for by the evolution of the baryonic content
within the progenitor halos.

\subsection{Evolution of the baryonic mass scatter}

As shown in the previous section, most of the scatter in the $z = 0.1$  $M_{*} - V_{\mathrm{max}}$ relation
is the result of scatter in the baryonic masses of the galaxies' progenitors,
most of which is due to variation in the baryonic mass fraction
of progenitor halos of the same mass. This
raises the question of what determines the baryon fraction.

In EAGLE, the baryonic mass within halos is primarily dependent on the halo mass.
Figure \ref{graph:baryonic_mass} shows the evolution of the distribution of baryonic masses as a function of $M_{\mathrm{200c, DMO}}$. (The results using $M_{\mathrm{200c}}$ from the reference simulation are very similar.)
The sample comprises halos with $M_{\mathrm{200c}} > 10^{11} \mathrm{M_{\odot}}$ for $z = 0.1$, 
$M_{\mathrm{200c}} > 10^{10.7} \mathrm{M_{\odot}}$ for $z = 1.0$, and $M_{\mathrm{200c}} > 10^{10.3} \mathrm{M_{\odot}}$ for $z = 2.0$. The masses of the latter
two redshifts are chosen to approximately encompass the masses of the halos hosting the main progenitors of the
$z = 0.1$ sample. The darkness of the shading is proportional to the log of the
number of halos in each bin.
The solid black line in each panel shows the baryonic mass that would be expected if each halo contained 
the cosmic baryon fraction times $M_{\mathrm{200c}}$.

At $z = 0.1$, the median value of $M_{\mathrm{bary}}$ as a function of $M_{\mathrm{200c, DMO}}$ is represented by a solid red line,
and the top and bottom deciles are shown with red dashed lines.
For high-mass halos ($M_{\mathrm{200c, DMO}} \gtrsim 10^{13} \mathrm{M_{\odot}}$), which are very low in number in EAGLE, the baryon fraction is close to
the cosmic value. However, for lower-mass halos, the mean baryon fraction is significantly lower. 

In the lower two panels, the median value of $M_{\mathrm{bary}}$ is shown with a solid cyan line, and the 
top and bottom deciles are
represented by dashed cyan lines. The deciles at $z = 0.1$ are replicated as red dotted lines. By comparing the top and
bottom deciles at $z = 0.1$
to those at $z = 1.0$ and $z= 2.0$, we see that for halos with $M_{\mathrm{200c, DMO}} \lesssim 10^{13} \mathrm{M_{\odot}}$, the
mean baryon fraction at fixed $M_{\mathrm{200c, DMO}}$ decreases with cosmic time and the scatter in the baryon fraction increases.

\begin{figure}
  \begin{center}
    \includegraphics[width=\columnwidth]{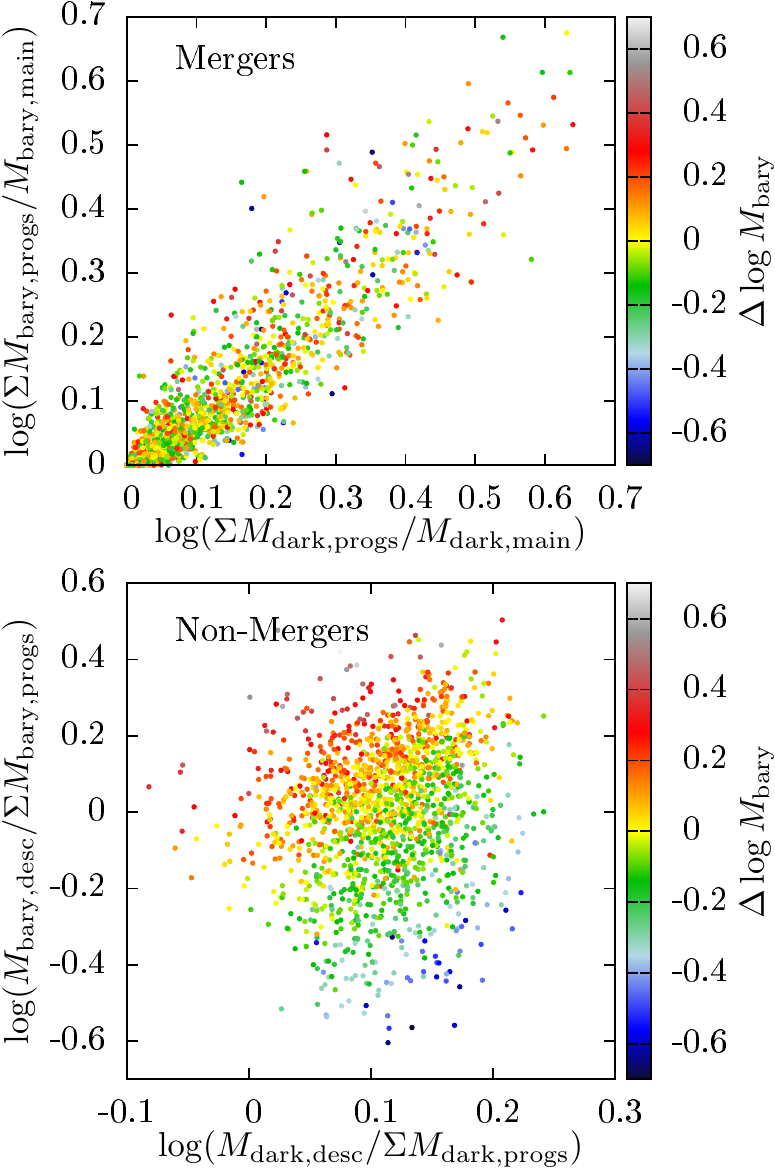}
    \caption{The influence of mergers and non-merger processes such as accretion and feedback on the evolution of 
$\Delta \log M_{\mathrm{bary}}$, the deviation of the baryonic mass of each halo
relative to the mean at fixed $M_{\mathrm{200c, DMO}}$ and $c_{\mathrm{DMO}}$ (see Eqn. \ref{delta_eq}).
The halo sample comprises central subhalos at  
$z = 1$ whose main progenitors at $z = 2$ 
are within 0.02 dex of the mean $M_{\mathrm{bary}}$ as a function of 
$M_{\mathrm{200c, DMO}}$ and $c_{\mathrm{DMO}}$ at $z = 2$. 
The color bar indicates $\Delta \log M_{\mathrm{bary}}$ of the descendants at $z = 1$ computed relative
to their halo properties at this redshift, 
showing that $\Delta \log M_{\mathrm{bary}}$ has scattered significantly to both larger and smaller values. 
\textit{Top Panel:} The growth in dark matter mass from mergers between $z = 2$
and $z = 1$ versus the growth in baryonic mass from mergers. 
The growth due to mergers is defined as the ratio of the sum of all the progenitor masses
to the mass of only the main progenitor. 
The dark matter mass is the total mass
in dark matter assigned to each FoF halo in the reference simulation.
The growth in dark matter and baryonic mass resulting from 
mergers correlates poorly with the final 
$\Delta \log M_{\mathrm{bary}}$ of the halo. 
\textit{Bottom Panel:} Same as the top panel, but for the change in mass not due to mergers (i.e. due to
accretion and feedback). The change in mass not due to mergers is defined as the ratio of the mass of the descendant at
$z = 1.0$ to the sum of all the progenitor masses at $z = 2.0$. The change in mass not due to mergers shows a 
far better correlation with $\Delta \log M_{\mathrm{bary}}$ of the descendant,
implying that feedback and gas accretion are the dominant contributors to the evolution of the baryon fraction. }
    \label{graph:mergers_env}
  \end{center}
\end{figure}

Halos in EAGLE undergo continuous evolution in the value of their baryonic mass
relative to their halo mass, so $\Delta \log M_{\mathrm{bary}}$ at low redshift ($z \approx 0$) is uncorrelated
with that at high redshift ($z \gtrsim 4$).
Evolution in $\Delta \log M_{\mathrm{bary}}$ results from change in both the 
dark matter mass and the baryonic mass of a halo, as well as the mean evolution of the sample of halos.
To determine the primary mechanism that sets the value of $\Delta \log M_{\mathrm{bary}}$,
we wish to compare the evolution of this value for each halo to the change
in the halo's dark matter and baryonic mass resulting from different physical processes ---
specifically, halo mergers versus non-merger processes such as accretion and feedback.

We select a sample consisting of halos at $z = 1.0$ with $M_{\mathrm{200c}} > 5\times10^{10} \mathrm{M_{\odot}}$,
such that the main progenitor of the central galaxy at $z = 2.0$ 
has a host halo with baryonic mass within 0.02 dex of the 
mean value for its $M_{\mathrm{200c, DMO}}$ and $c_{\mathrm{DMO}}$.
Stated differently, the $z = 2.0$ main progenitors have 
$|\Delta \log M_{\mathrm{bary}}| < 0.02$ relative to their $z = 2.0$ halo properties.
We then compute $\Delta \log M_{\mathrm{bary}} (z = 1.0)$, the deviation of $M_{\mathrm{bary}}$
from the mean at fixed $M_{\mathrm{200c, DMO}}$ and $c_{\mathrm{DMO}}$ at $z = 1.0$.
For the descendant halos at $z = 1.0$, 
the standard deviation of $\Delta \log M_{\mathrm{bary}}$ has increased to
0.19 dex, due to evolution in the baryonic
and dark matter masses of each halo since $z = 2.0$.

In order to consistently track the co-evolution of the dark matter and
baryonic masses, we use the total dark matter mass assigned
to each FoF halo in the reference simulation, denoted $M_{\mathrm{dark}}$. 
The baryonic mass $M_{\mathrm{bary}}$ is the same as described previously.
Because we are interested in the change in the \textit{total} dark matter and baryonic
mass within the halo, we consider the progenitors of the $z = 1.0$ halo
to be the host halos of the progenitors of both its central and satellite subhalos.
We note that due to the influence of baryonic physics, 
there are differences between the evolution of the FoF halo mass
in the reference simulation and that of
the corresponding halo in the DMO simulation. For the mass
range of the descendant sample considered here, the scatter between 
the FoF halo mass in the reference and DMO simulations
is $<0.06$ dex and decreases sharply with halo mass. We use
$M_{\mathrm{dark}}$ rather than the dark matter mass within $M_{\mathrm{200c}}$ 
because the former is more reflective of the accretion
of dark matter onto the halo.

In Figure \ref{graph:mergers_env} we show the change in halo 
dark matter and baryonic mass between $z = 2.0$ and $z = 1.0$, compared
to $\Delta \log M_{\mathrm{bary}} (z=1.0)$ for each halo.
The color of each point represents $\Delta \log M_{\mathrm{bary}}$ at $z= 1.0$,
which has evolved from a value of $\approx 0$ at $z = 2.0$.
The top panel of Figure \ref{graph:mergers_env} shows the mass growth due to mergers,
which we approximate as the ratio of the sum of the masses of all the progenitors
at $z = 2.0$ to the mass of the main progenitor\footnote{
This is an approximation because any mass accreted onto (or lost from) the
non-main progenitors after
$z = 2$ but before they merge with the main progenitor will not be considered
mass change from mergers but rather from non-mergers (second panel of Figure \ref{graph:mergers_env}).}: 
$\Sigma M_{\mathrm{progs}}/M_{\mathrm{main}}$.
Because the set of all progenitors includes the main progenitor, 
the mass change due to mergers is positive by definition. The vertical axis
shows the growth in the baryonic mass 
and the horizontal axis shows the growth in dark matter mass.

The growth in baryonic mass from mergers tends to
follow the growth in dark matter mass. Due to the low typical
mass of halos in our sample, the majority do not gain
a large amount of mass via mergers and are located in the
bottom left corner of the top panel of Figure \ref{graph:mergers_env}. However, 
even for those halos that experience a significant amount of growth from
mergers (primarily high-mass halos), 
$\Delta \log M_{\mathrm{bary}}$ of the descendant halo at $z = 1.0$ is effectively uncorrelated
with the change in either dark matter or baryonic mass due to mergers.
This suggests that
mergers are not the primary cause of change in $\Delta \log M_{\mathrm{bary}}$ over time. 

The lower panel of Figure \ref{graph:mergers_env} shows the change in dark matter
and baryonic mass due to non-merger processes, i.e. gas loss due to feedback and accretion
of dark matter and/or gas. The mass change due to non-merger processes is approximated
as the ratio of the mass of the $z = 1.0$ descendant to the sum of the masses of all its progenitors
at $z = 2.0$: $M_{\mathrm{desc}}/\Sigma M_{\mathrm{progs}}$.
The dark matter mass of the descendant is generally larger than the total dark matter
mass of the progenitors, but in some cases it can be smaller, perhaps because of ejection of matter
during mergers. The baryonic mass of the descendant, on the other hand, 
is frequently smaller than the sum of the baryonic masses of 
its progenitors. This indicates that feedback plays a very important role in changing the baryonic mass.

Furthermore, in contrast to the top panel,
$\Delta \log M_{\mathrm{bary}}$ at $z = 1.0$ correlates
clearly with the mass change caused by mechanisms other than mergers.
The change in $\Delta \log M_{\mathrm{bary}}$ (from a value of $\approx 0$
at $z = 2.0$) depends on the change in both the baryonic and dark matter mass
of each halo, as well as the mass evolution of the other halos.
The latter results from the fact that
$\Delta \log M_{\mathrm{bary}}$ is the deviation of $M_{\mathrm{bary}}$ relative to 
other halos of the same mass and concentration. 
As a result, although $\Delta \log M_{\mathrm{bary}}$ increases when the baryonic
mass of the halo increases and decreases when the dark matter mass increases,
the relationship in the lower panel of Figure 
\ref{graph:mergers_env} does not have a slope of one, 
because higher-mass halos have even more
baryonic mass than would be expected from a linear relation the two values,
as can be seen in Figure \ref{graph:baryonic_mass}.
For the same reason, halos in the lower panel 
of Figure \ref{graph:mergers_env} that remain at approximately
the same dark matter and baryonic mass tend to increase in $\Delta \log M_{\mathrm{bary}}$
between $z = 2.0$ and $z = 1.0$, because the baryonic mass
within halos at fixed halo mass decreases with redshift, as
can also be seen in Figure \ref{graph:baryonic_mass}.

Figure \ref{graph:mergers_env} shows that
the majority of the evolution in $\Delta \log M_{\mathrm{bary}}$ 
is attributable to change in the baryonic mass at fixed
values of accreted dark matter mass. We conclude that the evolution 
of $\Delta \log M_{\mathrm{bary}}$ over time is mainly due to inflow and outflow of gas
via feedback and smooth accretion, rather than mergers. 

\section{Discussion and Conclusions}

The EAGLE cosmological hydrodynamical simulation was previously used in 
\citet{matthee2017} and \citet{chaves2016}
to investigate the relationship between stellar mass $M_{*}$ and dark matter halo properties from
the dark matter-only (DMO) run of EAGLE, so as to determine the best parameter to use in halo abundance matching. Both
found that for central galaxies, 
the maximum circular velocity of the corresponding DMO halo, $V_{\mathrm{max, DMO}}$, correlates better with the stellar
mass than the DMO halo mass does, 
and that this relationship has a mass-dependent scatter that
is $\approx 0.2$ dex for halos with $M_{\mathrm{200c}} > 10^{11} \mathrm{M_{\odot}}$ at $z = 0.1$. \citet{matthee2017} investigated
whether the scatter in $M_{*}$ correlates with any other DMO halo properties, such as the 
halo half-mass assembly time, sphericity, spin, triaxiality, and environment, but found no additional
correlations.

In this paper, we have examined the source of the scatter in $M_{*}$ at fixed 
$V_{\mathrm{max, DMO}}$ for central galaxies by considering different baryonic (rather than dark matter) properties correlated
with the scatter. We used the same sample of central galaxies as \citet{matthee2017}, 
and the corresponding host halos from the DMO run of EAGLE.
Our main conclusion is that the scatter in $M_{*}$ at fixed $V_{\mathrm{max, DMO}}$ can
be traced primarily to the scatter in the baryon fraction of the host halos of
the galaxy progenitors.

In EAGLE, the baryonic mass of halos correlates primarily with the halo mass.
At high redshifts, the initial conditions are such that all halos have approximately the cosmic ratio
of baryons to dark matter. However, the mean baryonic mass at fixed halo mass for halos with 
$M_{\mathrm{200c}} \lesssim 10^{13} \mathrm{M_{\odot}}$ (which constitute the majority of our halo sample)
decreases with cosmic time, and the scatter in the baryonic mass at fixed halo mass increases,
as shown in Figure \ref{graph:baryonic_mass}.

The star formation rate of a halo's central galaxy 
depends on the central gas density, such that for an equal gas reservoir, 
a halo with a higher central density will produce more stars. 
Furthermore, a higher density implies a higher binding energy
and hence less efficient feedback for a fixed
rate of energy injection.
In addition, more concentrated halos tend to form earlier, allowing
more time for star formation to take place.
For these reasons, the 
stellar mass formed at fixed halo mass is higher for halos with 
higher concentrations, resulting in the stellar mass being better correlated with $V_{\mathrm{max, DMO}}$
than $M_{\mathrm{200c, DMO}}$. However, as described above, the baryon content of halos
of the same halo mass and concentration has a substantial scatter.
As a result, two halos with similar assembly histories 
but different baryonic mass fractions can produce descendant halos
with the same halo mass and concentration but 
significantly different stellar mass content.
We calculate the correlation of the scatter in the central stellar mass at fixed DMO halo
mass and concentration with the scatter in the baryonic masses of the galaxy progenitors.

The strongest correlation
between the scatter in $z = 0.1$ stellar mass and 
the scatter in the main progenitor baryonic mass
is achieved at $z \approx 0.9$, where the latter is able to account
for $67\%$ of the variance in the $z = 0.1$ $M_{*}-V_{\mathrm{max, DMO}}$ 
relation for halos with $M_{\mathrm{200c}} > 10^{11} \mathrm{M_{\odot}}$ (Figure \ref{graph:redshift_correlation},
top panel). 
The correlation with the sum of the baryonic masses of all the progenitors is slightly better, peaking for progenitors at 
$z \approx 1.1$, which account for $75\%$ of the variance
 in the $z = 0.1$ $M_{*}-V_{\mathrm{max, DMO}}$ relation.
Similar trends are seen in the lower panels of Figure \ref{graph:redshift_correlation} 
for samples of central galaxies at $z = 0.5$ and $z = 1.0$
having halo masses greater than $10^{11} \mathrm{M_{\odot}}$, with the location of the peak correlation
for the sum of the baryonic masses of all the progenitors 
shifted to $z \approx 1.5$ and $z \approx 2.0$, respectively. 

The peak strength of the correlation between the scatter in the $z = 0.1$ stellar mass
and that of the progenitor baryonic masses also depends on the descendant halo mass,
because higher-mass halos and their central galaxies 
gain more mass from mergers and have more stochastic growth histories.
This can be seen in Figure \ref{graph:mass_bins}.
The peak correlation between the scatter of the descendant stellar mass and 
that of the sum of the baryonic masses of all the progenitors
 is $77\%$ for halos with $10^{11.0} \mathrm{M_{\odot}} < M_{\mathrm{200c, DMO}} < 10^{11.4} \mathrm{M_{\odot}}$,
but $63\%$ for those with $M_{\mathrm{200c, DMO}} > 10^{12.2} \mathrm{M_{\odot}}$.
The peak correlation occurs at $z \approx 1$ regardless of the halo mass. 
In contrast, the redshift of peak correlation between 
the scatter in the baryonic mass of the main progenitor 
and the scatter in the descendant stellar mass does vary with halo mass,
since higher-mass halos obtain a larger fraction of their mass through late-time mergers.
The strength of the peak correlation for the main progenitor also varies more with halo mass. 
For $10^{11.0} \mathrm{M_{\odot}} < M_{\mathrm{200c, DMO}} < 10^{11.4} \mathrm{M_{\odot}}$,
the correlation between the scatter in the $z = 0.1$ stellar mass and main progenitor
baryonic mass peaks at $z \approx 0.9$, where it has a value of $72\%$,
while for $M_{\mathrm{200c, DMO}} > 10^{12.2} \mathrm{M_{\odot}}$, the correlation peaks at $z \approx 0.5$
with a value of $50\%$.

The scatter in the baryonic mass within halos also produces an inverse correlation
between the central galaxy's stellar mass and stellar population age at fixed DMO halo mass and concentration, shown
in the top panel of Figure \ref{graph:delta_age}.
The halos with more massive central galaxies at $z = 0.1$ are those that 
had a larger amount of recent star formation due to their larger baryon reservoir, 
causing their central galaxies to be more massive and younger.

Finally, we determined that non-merger processes, such
as gas accretion and feedback, are what primarily set
the baryonic mass within halos. 
The complex and stochastic nature of feedback
likely explains the lack of significant correlation
with the DMO halo properties examined in \citet{matthee2017}.
In a companion paper (Kulier et al. 2018, in prep), we describe in detail
the origin of variations in feedback strength for different halo mass ranges
and timescales and its correlates.

\section*{Acknowledgements}

The authors would like to thank Jorryt Matthee for very useful comments
on the first draft of the paper, and Claudia Lagos
for her suggestions regarding this research.
We would also like to thank the anonymous 
referee for their helpful suggestions.

This work was supported by the Netherlands 
Organisation for Scientific Research (NWO), through VICI grant 639.043.409
and VENI grant 639.041.749,
as well as by the Science and Technology Facilities Council [ST/P000541/1]. 
AK acknowledges support from CONICYT-Chile grant FONDECYT Postdoctorado 3160574.
RAC is a Royal Society University Research Fellow.

This work used the DiRAC Data Centric system at Durham University, operated by the
Institute for Computational Cosmology on behalf of the STFC DiRAC HPC Facility
(www.dirac.ac.uk). This equipment was funded by BIS National E-infrastructure
capital grant ST/K00042X/1, STFC capital grant ST/H008519/1, and STFC DiRAC
Operations grant ST/K003267/1 and Durham University. DiRAC is part of the National
E-Infrastructure.

\bibliographystyle{mn2e}

\label{lastpage}

\end{document}